\documentclass[%
reprint,
 amsmath,amssymb,
 aps,
 prl,
]{revtex4-2}

\usepackage{graphicx}% Include figure files
\usepackage{dcolumn}% Align table columns on decimal point
\usepackage{bm}% bold math
\usepackage{xr-hyper} % crossreference to SI
\usepackage{hyperref}% add hypertext capabilities
\usepackage{xcolor} % colored text to exchange comments.
\usepackage{textcomp, gensymb}

\makeatletter

\begin{document}

\title[Sliding ferroelectricity in a bulk misfit layer compound (PbS)$_{1.18}$VS$_2$]{Sliding ferroelectricity in a bulk misfit layer compound (PbS)$_{1.18}$VS$_2$}% Force line breaks with \\

\author{Cinthia Antunes Corr\^{e}a}
\affiliation{FZU - Institute of Physics of the Czech Academy of Sciences, Na Slovance 2, Praha 8, CZ-18 221, Czech Republic}
\affiliation{Department of Physics of Materials, Faculty of Mathematics and Physics, Charles University, Ke Karlovu 3, CZ-121 16, Prague 2, Czech Republic}

\author{Ji\v{r}\'{i} Voln\'{y}}
\author{Kate\v{r}ina Tetalov\'{a}}
\author{Kl\'{a}ra Uhl\'{i}\v{r}ov\'{a}}
\affiliation{Department of Condensed Matter Physics, Faculty of Mathematics and Physics, Charles University, Ke Karlovu 3, CZ-121 16, Prague 2, Czech Republic}

\author{V\'{a}clav Pet\v{r}\'{i}\v{c}ek}
\affiliation{FZU - Institute of Physics of the Czech Academy of Sciences, Na Slovance 2, Praha 8, CZ-18 221, Czech Republic}

\author{Martin Vondr\'{a}\v{c}ek}
\author{Jan Honolka}
\affiliation{FZU - Institute of Physics of the Czech Academy of Sciences, Na Slovance 2, Praha 8, CZ-18 221, Czech Republic}

\author{Tim Verhagen }
\email{verhagen@fzu.cz}
\affiliation{FZU - Institute of Physics of the Czech Academy of Sciences, Na Slovance 2, Praha 8, CZ-18 221, Czech Republic}
\affiliation{Institute of Physics, Faculty of Mathematics and Physics, Charles University, Ke Karlovu 3, CZ-121 16, Prague 2, Czech Republic}

\begin{abstract}
Twisted heterostructures of two-dimensional crystals can create a moiré landscape, which can change the properties of it's parent crystals. However,the reproducibility of manual stacking is far from perfect. Here, the alternated stacking of post-transition metal monochalcogenides and transition metal dichalcogenides in misfit layer compound crystals is used as a moir\'{e} generator. Using X-ray diffraction, the presence of twins with a well-defined small twist angle between them is shown. Due to the twist, the surface electrical potential from the induced ferroelectricity is observed using scanning probe microscopy and electron microscopy.
\end{abstract}

\maketitle

The group of possible two-dimensional (2D) ferroelectric materials has been significantly expanded by the recent theoretical prediction~\cite{Li2017, Yang2018, Yang2023, atri2023} and experimental confirmation~\cite{Vizner2021, Yasuda2021, Weston2022, Rogee2022, Woods2021, garciaruiz2023} of sliding ferroelectricity. For certain stacking configurations in bi- and multilayer 2D materials, the inversion and/or mirror symmetry is broken and thereby an in- and/or out-of-plane polarization are induced. Due to the weak interlayer interaction in van der Waals multilayers, the polarization can be reversed via in-plane lattice sliding. 

Experimentally, a sliding ferroelectric material can be created by manual stacking of 2D van der Waals materials on top of each other with a defined twist angle $\theta$~\cite{Carr2018, Gargiulo2018, Rosenberger2020, Weston2020}. Unfortunately, manual stacking can create difficulties with device reproducibility~\cite{Lau2022}. Sliding ferroelectricity can also be observed in bulk van der Waals crystals, as long as there is no inversion symmetry between the layers, as was recently shown for the amphidynamic crystal (15-crown-5)Cd$_3$Cl$_6$~\cite{Miao2022}. In the case of transition metal dichalcogenides (TMD) crystals, the centrosymmetric 2H-form is the most common stable form and rarely a 3R-stacked bilayer can be found~\cite{Weston2020}. Nevertheless, bulk crystals with a broken inversion symmetry can be grown, such as 3R-MoS$_2$~\cite{Suzuki2014} or graphene polytypes with stacking faults or twins~\cite{atri2023, garciaruiz2023}.  Also in van der Waals epitaxy of 2D materials on (quasi)-2D substrates, the formation of stacking faults and twins is often observed,  due to very similar binding energies for the different stacking  configurations~\cite{Mortelmans2021}.

In this work, instead of manually stacking a van der Waals heterostructure with a certain twist angle, we searched for a class of 2D materials where twins having a small twist angle with respect to each other can be easily induced during the growth. A member of the TMD class are the misfit layer compounds (MLCs) with the general formula \textit{(MX)$_{1+m}$(TX$_2$)$_n$}, where \textit{M} = Sn, Pb, Sb, Bi, or a rare earth element; \textit{T} = Ti, V, Cr, Nb, or Ta; \textit{X} = S, or Se; 0.08 $<$ \textit{m} $<$ 0.28; and \textit{n} = 1, 2, or 3~\cite{Wiegers1996}. MLCs, characterized by the alternating stacking of a TMD layer and a post-transition metal chalcogenide monolayer (TMM), are the ideal class of 2D materials which should favor the creation of twins. During the growth, each next layer of 2D TMD(TMM) is grown on a TMM(TMD) substrate layer, which should be similar as with van der Waals epitaxy of thin films, result in the creation of twins.

Using a combination of single crystal X-ray diffraction analysis (XRD), scanning electron microscopy (SEM), photoemission electron microscopy (PEEM) and scanning probe microscopy (SPM), we show that within (PbS)$_{1.18}$VS$_2$ bulk crystals, twins are present. The twist angle between different twins is small and that twins with 3R-stacking show the characteristic properties of a sliding ferroelectric material.

The crystal structure of a MLC (also called composite) is formed by two or more distinct columns or layers, where each of them is called a subsystem. Two subsystems $\textbf{A}^{(1)}$ and $\textbf{A}^{(2)}$ are related by an interlattice matrix $\sigma$ via $\textbf{A}^{(2)}=\sigma\textbf{A}^{(1)}$~\cite{Vaclav1991, Cisarova1993}. The interaction between the subsystems corresponds to a perturbing potential that modulates the columns or layers and generates satellite reflections on the diffraction pattern of the composite. The intensities of the satellite reflections depend on the modulation. Since the diffraction pattern of a composite contains the contribution of each subsystem and the satellites, at least four indices are necessary to index it. 

\begin{figure}
 \includegraphics[width=\columnwidth] {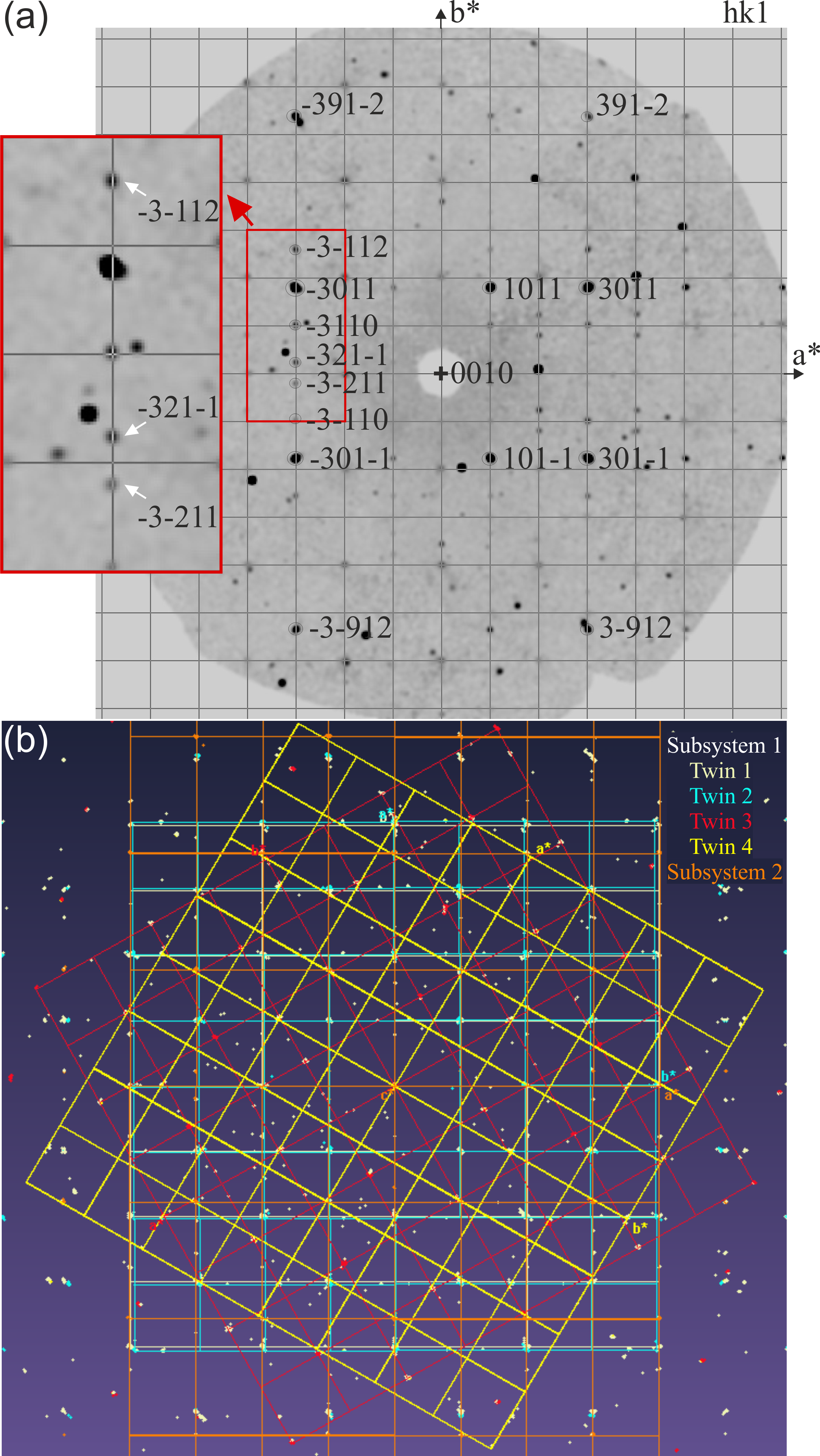}
  \caption{Reciprocal space map showing satellites and twins observed by XRD. (a) Section $hk1$ of (PbS)$_{1.18}$VS$_2$. In the inset, the position of several reflections with lower intensity is highlighted using white arrows. (b) Reciprocal space map from CrysAlisPro~\cite{crysalis}: from subsystem 1, twin 1 is shown in light yellow, twin 2 in light blue, twin 3 in red, and twin 4 in yellow. The second subsystem is shown in orange.}
  \label{fgr:fig_1}
\end{figure}

(PbS)$_{1.18}$VS$_2$ crystals were grown using chemical vapor transport according to a modified recipe of Gotoh et al.~\cite{Gotoh1990, suppl}. They crystallize as a layer composite structure (see Figs.~\ref{fgr:fig_s1}~\textendash~\ref{fgr:fig_s3} in the Supplemental Material~\cite{suppl}), where subsystems 1 and 2 are PbS and VS$_2$, respectively. Subsystems 1 and 2 have parallel sublattice parameters $\textbf{a}$ and $\textbf{c}$, while $\textbf{b}$ is the incommensurate axis. Using the superspace formalism, indices $\textbf{a}^*$, $\textbf{b}_1^*$, $\textbf{c}^*$, and $\textbf{b}^*_2$ \cite{Cisarova1993} are necessary to index both reciprocal sublattices of (PbS)$_{1.18}$VS$_2$, where $\textbf{b}_1^*$ is the modulation vector of sublattice 2, and $\textbf{b}_2^*$ is the modulation vector of sublattice 1.

A diffraction vector $\textbf{Q}$ is given by $\textbf{Q}=h\textbf{a}^*+k\textbf{b}_1^*+l\textbf{c}^*+m\textbf{b}^*_2$, where $hklm$ are the reflection indices containing information from both subsystems;

\begin{itemize}
    \item $k=m=0$, main reflections common for both subsystems;
    \item $k\neq0$ $m=0$, main reflections for the first subsystem, $k^{th}$ order satellites for the second subsystem;
    \item $k=0$, $m\neq0$, main reflection for the second subsystem, $m^{th}$ order satellites;
    \item $k\neq0$, $m\neq0$, pure satellite reflection of $k^{th}$ order and $m^{th}$ order for the first and second subsystem, respectively.
\end{itemize}

\begin{figure*}
 \includegraphics[width=2\columnwidth] {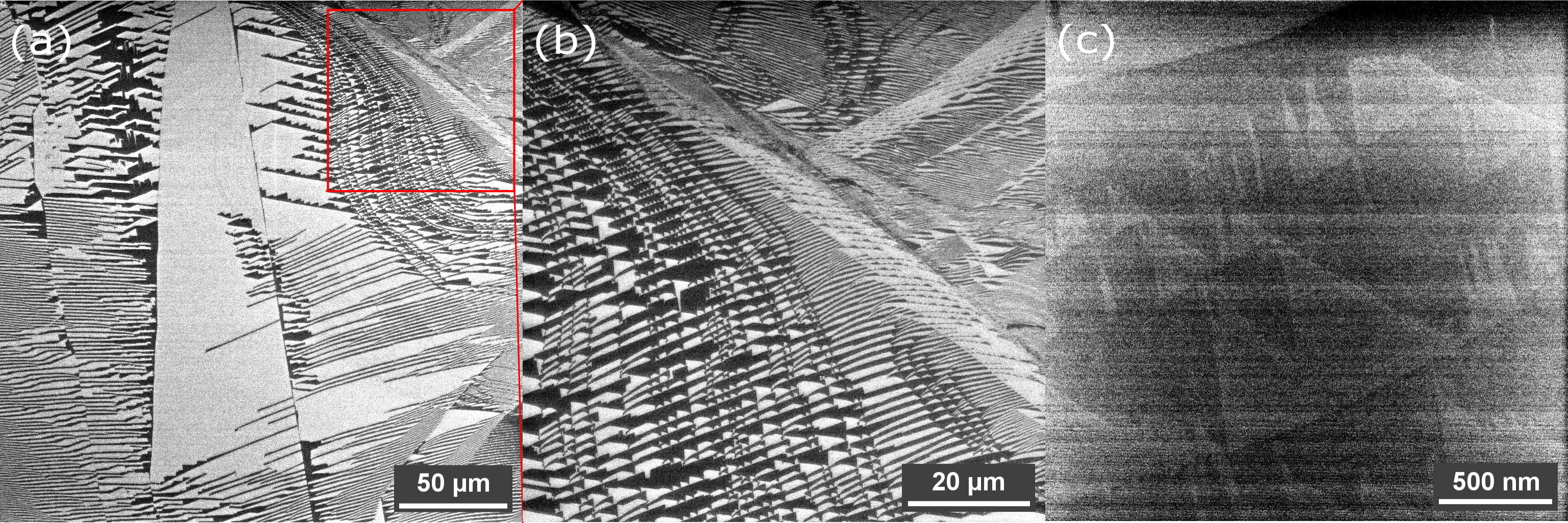}
  \caption{(a) SE image of a cleaved (PbS)$_{1.18}$VS$_2$ crystal, where a complex domain structure is visible with domain sizes up to hundreds of micrometers. A close-up of the top-right corner of the panel (a) is shown in panel (b), with domain sizes present in the micrometer range. In panel (c), a further zoom-in on a different location shows that also on the sub-micrometer size, a variety of domains exists.}
  \label{fgr:fig_3}
\end{figure*}
 
In a previous attempt to elucidate the crystal structure of (PbS)$_{1.18}$VS$_2$, Gotoh et al.~\cite{Gotoh1990, Onoda1990} observed satellites on electron diffraction photographs, but not by single crystal XRD. Hence, the structure was analyzed using the superspace formalism without observed satellites.

In our case, not only satellites were observed by single crystal X-ray diffraction, but also twins were present in the data, as shown in the reconstructed reciprocal space section $(hk1)$ (Fig.~\ref{fgr:fig_1}). Using a monoclinic unit cell to index both subsystems, the $a$ and $c$ parameters are common to both subsystems: subsystem 1 (PbS) has unit cell parameters $a=5.7062(5)$, $b=5.7674(7)$, $c=23.691(2)$ ~\AA, and $\beta=94.878(8)\degree$. Subsystem 2 (VS$_2$) has $b=3.204$ ~\AA. Figure~\ref{fgr:fig_1}(a) contains reflections belonging to subsystems 1 and 2, weak satellites, and several reflections that are not indexed. The extra reflections can be indexed, if other twins are included during the data integration. Since the two subsystems have either a distorted cubic or a hexagonal subcell, there are several possible configurations of layer organization, which allows for a range of twin orientations. Figure~\ref{fgr:fig_1}(b) shows the reciprocal space map containing four possible twins from the PbS subsystem: twin 1 has 30.8$\%$ of indexation, twin 2 has 26.6$\%$ of indexation, twin 3 has 24.5$\%$ of indexed peaks, and twin 4 has 4.8$\%$ of indexed peaks. The second subsystem is shown in orange and has 19.7$\%$ of indexed peaks. Each twin has part of the indexed peaks overlapped with other components. Data integrated for the structure analysis~\cite{crystal_structure} includes three twins, which can be seen in Fig.~\ref{fgr:fig_s3}(a) in the Supplemental Material~\cite{suppl}. The first twin is the one with the highest percentage of indexed reflections and has the main reflections from subsystems 1 and 2, plus satellites (Fig.~\ref{fgr:fig_s3}(b) in the Supplemental Material~\cite{suppl}). Even though the relation between twins 2, 3, and 4 to twin 1 seems to be a rotation of approximately 180$\degree$, 30$\degree$, and 60$\degree$ around $c^*$, respectively (see Fig.~\ref{fgr:fig_1}(b)), the real rotation is along all the three axes $a^*$, $b^*$, and $c^*$.

From a crystallographic point of view, a variety of twin angles can be expected, but the dominant twist angles here, have a preference angle of approximately 0 and $\pm$180$\degree$ with respect to the twin with the highest indexation. Table.~\ref{tab:tab_s1} in the Supplemental Material~\cite{suppl}  shows this effect in more detail. This is probably no coincidence, as the difference between the twist angle of two twins being approximately 0$\degree$ twins would result in an 3R-type alignment and 60 or 180$\degree$ twins would result in an 2H-type alignment~\cite{Rosenberger2020} of the VS$_2$-layer, respectively. This indicates that these two configurations are local energy minima, similar as has been observed in exfoliated van der Waals heterostructures~\cite{Rosenberger2020}.

To investigate the possible presence of sliding ferroelectricity between the different twins, secondary electrons imaging (SE) in a SEM was used to map the surface electrical potential and spatial variations of topography~\cite{Andersen2021, LEBIHAN1972, Hunnestad2020}. In Fig.~\ref{fgr:fig_3}, SE images of the (PbS)$_{1.18}$VS$_2$ surface cleaved in ambient conditions are shown. Clearly visible is the presence of a complex domain structure, where the shape of the domains is either preferentially aligned stripes or triangular-like. Similar domain shapes were observed before in exfoliated TMD-based van der Waals heterostructures, which were stacked on top of each other with a small twist angle between the layers. Those domains were shown to be ferroelectric~\cite{Vizner2021, Yasuda2021, Weston2022, Rogee2022}. The triangular domain indicates that the layers in those areas have a 3R-type alignment, and the triangular moir\'{e} is induced by the distorted hexagonal VS$_2$ layer. SE imaging of a 2H-type alignment is shown in Fig.~\ref{fgr:fig_s6} in the Supplemental Material~\cite{suppl}.

The period $l$ of the domain network is for very small twist angles $\theta$ equal to $a/\theta$, where $a$ is the lateral lattice constant of VS$_2$~\cite{Weston2022}. The largest observed triangular shape domains with lateral dimensions around 30~$\mu$m suggest that the smallest twist angles present between the hexagonal lattices of the VS$_2$ layers are of the order 10$^{-4}$ rad. 

\begin{figure*}
 \includegraphics[width=2\columnwidth] {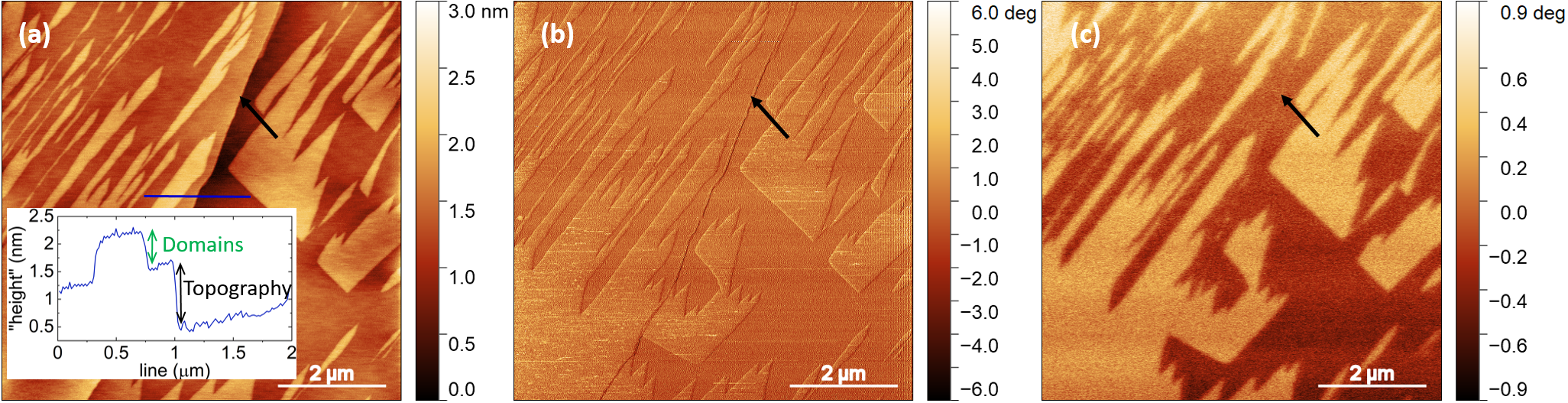}
  \caption{Topography (a) and TM-AFM phase (b) of a cleaved (PbS)$_{1.18}$VS$_2$ crystal, where the domain landscape is imprinted onto the topography. The black arrow in each panel indicates the topography step edge in the image. The inset in panel (a) shows a height profile along the blue line (main AFM image), indicating a topography step edge and an imprinted domain. (c) EFM of the charge polarization of the same area shown in panel (a).}
  \label{fgr:fig_4}
\end{figure*}

As recently shown, the ferroelectric domains in van der Waals moir\'{e} materials can be straightforwardly imaged using electric force microscopy (EFM)~\cite{Woods2021,Moore2021}. Even in standard tapping mode atomic force microscopy (TM-AFM), the phase and even topography can be used for imaging moir\'{e}  interlayer modulation of the van der Waals potential~\cite{Chiodini2022}. Our TM-AFM and EFM measurements were performed on freshly cleaved single crystals. Like the SEM images presented in Fig.~\ref{fgr:fig_3}, domain structures with similar shapes and sizes were found, depending on the position at the sample surface. In Fig.~\ref{fgr:fig_4}, the topography and phase of the TM-AFM and EFM's phase are shown. When Fig.~\ref{fgr:fig_4}(a) and (b) are compared, two different features can be seen in both the topography and TM-AFM phase contrast; the terrace step going through the whole image, marked by the arrow, and the domains. In the topography image, the features appear as approximately 1~nm high steps. In the TM-AFM phase image, the domains induce a constant phase shift,  resulting from different energy dissipation due to the modulation of van der Waals forces in the moir\'{e} structures~\cite{Chiodini2022} while the terraces are visible only as lines along the step edges. 

The domains are also visible in the EFM phase image in Fig.~\ref{fgr:fig_4}(c) (lift height 15~nm, zero bias voltage). The domains show a large contrast,  while the terrace edge is barely visible. When a bias voltage is applied, the phase difference between both domains becomes larger, and the contrast can be switched when an opposite bias voltage is applied (not shown).  This effect is similar to measurements on der Waals heterostructures~\cite{Chiodini2022, Weston2022, Yasuda2021, Woods2021}. However, in our case, the domains are much more pronounced in the topography than reported. This is probably due to the samples' bulk character and surface quality.

Energy-filtered PEEM imaging can provide spatial information on the elemental composition, but also on changes in the work function caused by the different alignment of the electric polarization, as already shown in the early 1970s by Le Bihan~\cite{Bihan1970, Bihan1972} and Morlon et al.~\cite{Morlon1970} for ferroelectric domains in BaTiO$_3$.

\begin{figure*}
 \includegraphics[width=2\columnwidth] {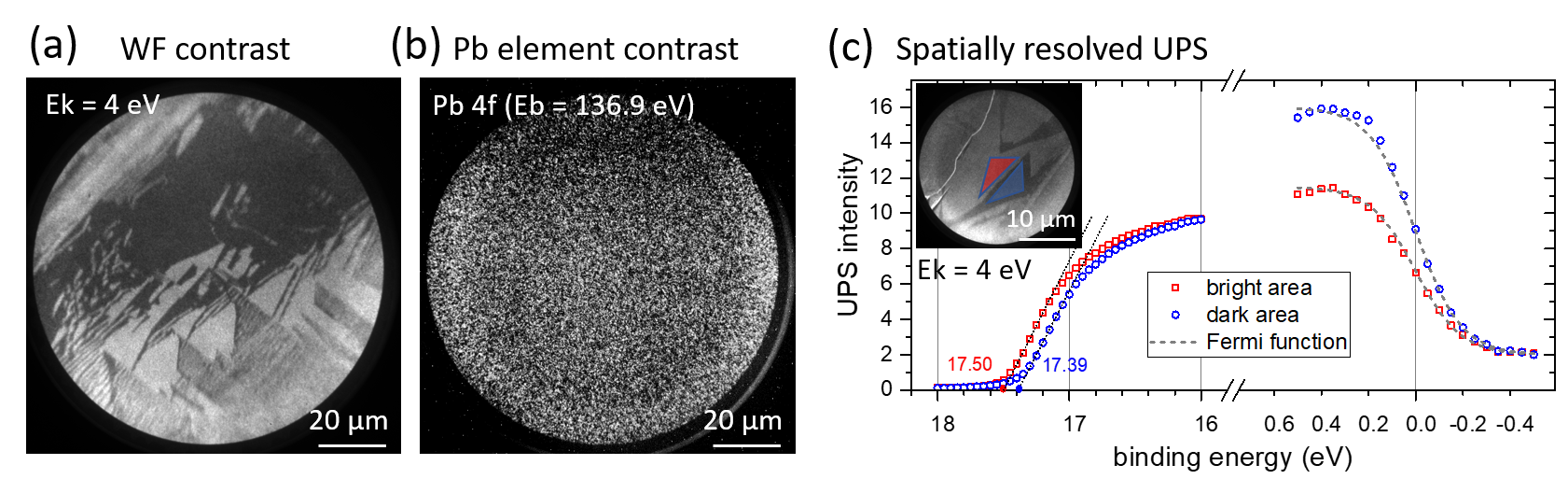}
  \caption{(a) Energy filtered PEEM image (kinetic energy $E_k = 4$~eV) of an in-situ cleaved (PbS)$_{1.18}$VS$_2$ crystal. (b) Element map of the same area taken using energy-filtered PEEM of Pb 4f (electron binding energy $E_b = 136.9$~eV with an analyser energy resolution of 0.4~eV).
  (c) Spatially resolved UPS data extracted from bright and dark domains marked in red and blue in the inset PEEM image at $E_k = 4$~eV. In the UPS mode the analyser energy resolution is 100~meV. The Fermi edge is located at $E_b = 0$~eV in both cases but the cut-off energy of secondary electrons have different values 17.50 and 17.39 eV for bright and dark areas, respectively.}
  \label{fgr:fig_2}
\end{figure*}

In Fig.~\ref{fgr:fig_2}(a), an energy-filtered PEEM image is shown acquired using a Hg lamp and an analyzer kinetic energy setting of 4~eV. In this mode, the image is dominated by secondary electrons and reflects differences in the work function. The work function contrast reveals a domain structure similar to what was also observed using SEM and SPM. To verify if these domains are not caused by alternations of either a PbS or VS$_2$ termination of the crystal, an elemental map of the same area was measured using energy-filtered PEEM of the Pb 4f core spectrum  (see Fig.~\ref{fgr:fig_s4} in the Supplemental Material~\cite{suppl} for the complete XPS spectrum). 
As the intensity of the PEEM imaging is very sensitive to the stacking of the final layers of the crystal, a different termination of the sample should result in the appearance of PbS and VS$_2$ domains. As shown in Fig.~\ref{fgr:fig_2}(b), no contrast is visible, so it can be concluded that the termination of the whole area is the same.

Subsequently, spatially resolved ultraviolet photoelectron spectroscopy (UPS) was used to measure the work function of both domains. In Fig.~\ref{fgr:fig_2}(c), the UPS spectra of the two polarized regions are shown for low binding energies, [-0.5 - +0.5~eV], and high binding energies, [16 - 18~eV]. The low energy window reflects the density of states at the valence band maximum, whereas the high energy window reflects the drop in intensity due to the cut-off energy of secondary electrons. Comparing the data of both domains, it is visible that the Fermi edge remains at the same position. The fit by a Fermi function, dashed lines in Fig.~\ref{fgr:fig_2}(c), gives a Fermi energy position $E_b = (0.00 \pm 0.02)$~eV. In contrast, the secondary electron energy cut-off around 17.5~eV, derived from the tangential fit, is downshifted by 0.11~eV between the two domains. This indicates that the change in the UPS spectra is due to a different work function of both domains.

In the following part, we discuss the consequences of the observed twins and domains in the (PbS)$_{1.18}$VS$_2$ crystals. In contrast to `traditional' electrically isolating ferroelectrics, (PbS)$_{1.18}$VS$_2$ crystals are semiconducting~\cite{Wiegers1996}, according to electric transport measurements, or even metallic, according to UPS or UV-VIS absorption spectroscopy~\cite{Sourisseau1995}. The comparably large conductivity makes the `traditional' electrical characterization used to identify ferroelectric materials and to switch the ferroelectric domains very challenging. Nevertheless, here we show using the combination of the structural data obtained with single crystal XRD and the observed domains obtained using various technique that sliding ferroelectricity is present in bulk MLC crystals . 

Generally, the crystal structure refinement of MLCs using single crystal XRD is challenging~\cite{Ng2022}. However, moir\'{e} patterns in van der Waals heterostructures, which work as a magnifying glass~\cite{Cosma2014}, provide essential information for the structural refinement of the very small twist angle and strain variations between the different layers. 

The presence of moir\'{e} structures in (PbS)$_{1.18}$VS$_2$ make this class of materials a possible next platform to explore the physics of strong electronic correlations and possible non-trivial band topology. A first hint of such strongly correlated physics can be found in low-temperature studies which were performed in the past on the MLC LaS$_{1.17}$VS$_2$, which can be either electron or hole-doped using Pb and Sr~\cite{Ino2004, Yasui1995}.

To conclude, sliding ferroelectricity has been observed in bulk (PbS)$_{1.18}$VS$_2$ crystals. The mutual interaction between the PbS and VS$_2$ subsystems is the key mechanism, which introduces well-defined groups of twins in the (PbS)$_{1.18}$VS$_2$ bulk crystal, where the stacking of the twins determines if there is locally a 2H- or 3R-type alignment. The possibility to steer interaction-driven twist angles in bottom-up growth can provide a new pathway to create large-scale moir\'{e} landscapes, twintronics, both in thin films and bulk crystals and might remove the problems with manual stacked devices~\cite{Lau2022}.

% to add references in the SI
\nocite{Gotoh1990, LEBIHAN1972, Hunnestad2020, Andersen2021, esprit, jana2006, crysalis, superflip, vesta, Ettema1993, Brandt2003, Hossain2018, Zhang2017, Hossain2021, Silversmit2004, Goehler2022, Goehler2018, LEIRO1998, Fong2006, Baniecki2009, Krug2010}

 We thank Peter Min\'{a}rik for the FIB lamella preparation and Jan Michali\v{c}ka for the support during the HRTEM measurement. The authors acknowledge the support provided by the Czech Science Foundation (project No.~22-04816S) and from the Czech  Ministry of  Education,  Youth, and  Sports  (MEYS)  project  SOLID21  (Project No.~CZ.02.1.01/0.0/0.0/16$_{-}$019/0000760). Single-crystal growth and characterization were performed in MGML, within the program of Czech Research Infrastructures (project no.~LM2023065). CzechNanoLab (project no.~LM2023051) funded by MEYS CR is acknowledged for the financial support of the TEM measurements at CEITEC Nano Research Infrastructure.

% bbl
%


\begin{thebibliography}{55}%
\makeatletter
\providecommand \@ifxundefined [1]{%
 \@ifx{#1\undefined}
}%
\providecommand \@ifnum [1]{%
 \ifnum #1\expandafter \@firstoftwo
 \else \expandafter \@secondoftwo
 \fi
}%
\providecommand \@ifx [1]{%
 \ifx #1\expandafter \@firstoftwo
 \else \expandafter \@secondoftwo
 \fi
}%
\providecommand \natexlab [1]{#1}%
\providecommand \enquote  [1]{``#1''}%
\providecommand \bibnamefont  [1]{#1}%
\providecommand \bibfnamefont [1]{#1}%
\providecommand \citenamefont [1]{#1}%
\providecommand \href@noop [0]{\@secondoftwo}%
\providecommand \href [0]{\begingroup \@sanitize@url \@href}%
\providecommand \@href[1]{\@@startlink{#1}\@@href}%
\providecommand \@@href[1]{\endgroup#1\@@endlink}%
\providecommand \@sanitize@url [0]{\catcode `\\12\catcode `\$12\catcode
  `\&12\catcode `\#12\catcode `\^12\catcode `\_12\catcode `\%12\relax}%
\providecommand \@@startlink[1]{}%
\providecommand \@@endlink[0]{}%
\providecommand \url  [0]{\begingroup\@sanitize@url \@url }%
\providecommand \@url [1]{\endgroup\@href {#1}{\urlprefix }}%
\providecommand \urlprefix  [0]{URL }%
\providecommand \Eprint [0]{\href }%
\providecommand \doibase [0]{http://dx.doi.org/}%
\providecommand \selectlanguage [0]{\@gobble}%
\providecommand \bibinfo  [0]{\@secondoftwo}%
\providecommand \bibfield  [0]{\@secondoftwo}%
\providecommand \translation [1]{[#1]}%
\providecommand \BibitemOpen [0]{}%
\providecommand \bibitemStop [0]{}%
\providecommand \bibitemNoStop [0]{.\EOS\space}%
\providecommand \EOS [0]{\spacefactor3000\relax}%
\providecommand \BibitemShut  [1]{\csname bibitem#1\endcsname}%
\let\auto@bib@innerbib\@empty
%</preamble>
\bibitem [{\citenamefont {Li}\ and\ \citenamefont {Wu}(2017)}]{Li2017}%
  \BibitemOpen
  \bibfield  {author} {\bibinfo {author} {\bibfnamefont {L.}~\bibnamefont
  {Li}}\ and\ \bibinfo {author} {\bibfnamefont {M.}~\bibnamefont {Wu}},\ }\href
  {\doibase 10.1021/acsnano.7b02756} {\bibfield  {journal} {\bibinfo  {journal}
  {ACS Nano}\ }\textbf {\bibinfo {volume} {11}},\ \bibinfo {pages} {6382}
  (\bibinfo {year} {2017})}\BibitemShut {NoStop}%
\bibitem [{\citenamefont {Yang}\ \emph {et~al.}(2018)\citenamefont {Yang},
  \citenamefont {Wu},\ and\ \citenamefont {Li}}]{Yang2018}%
  \BibitemOpen
  \bibfield  {author} {\bibinfo {author} {\bibfnamefont {Q.}~\bibnamefont
  {Yang}}, \bibinfo {author} {\bibfnamefont {M.}~\bibnamefont {Wu}}, \ and\
  \bibinfo {author} {\bibfnamefont {J.}~\bibnamefont {Li}},\ }\href {\doibase
  10.1021/acs.jpclett.8b03654} {\bibfield  {journal} {\bibinfo  {journal} {J.
  Phys. Chem. Lett.}\ }\textbf {\bibinfo {volume} {9}},\ \bibinfo {pages}
  {7160} (\bibinfo {year} {2018})}\BibitemShut {NoStop}%
\bibitem [{\citenamefont {Yang}\ and\ \citenamefont {Wu}(2023)}]{Yang2023}%
  \BibitemOpen
  \bibfield  {author} {\bibinfo {author} {\bibfnamefont {L.}~\bibnamefont
  {Yang}}\ and\ \bibinfo {author} {\bibfnamefont {M.}~\bibnamefont {Wu}},\
  }\href {\doibase 10.1002/adfm.202301105} {\bibfield  {journal} {\bibinfo
  {journal} {Adv. Funct. Mater.}\ }\textbf {\bibinfo {volume} {n/a}},\ \bibinfo
  {pages} {2301105} (\bibinfo {year} {2023})}\BibitemShut {NoStop}%
\bibitem [{\citenamefont {Atri}\ \emph {et~al.}(2023)\citenamefont {Atri},
  \citenamefont {Cao}, \citenamefont {Alon}, \citenamefont {Roy}, \citenamefont
  {Stern}, \citenamefont {Falko}, \citenamefont {Goldstein}, \citenamefont
  {Kronik}, \citenamefont {Urbakh}, \citenamefont {Hod},\ and\ \citenamefont
  {Shalom}}]{atri2023}%
  \BibitemOpen
  \bibfield  {author} {\bibinfo {author} {\bibfnamefont {S.~S.}\ \bibnamefont
  {Atri}}, \bibinfo {author} {\bibfnamefont {W.}~\bibnamefont {Cao}}, \bibinfo
  {author} {\bibfnamefont {B.}~\bibnamefont {Alon}}, \bibinfo {author}
  {\bibfnamefont {N.}~\bibnamefont {Roy}}, \bibinfo {author} {\bibfnamefont
  {M.~V.}\ \bibnamefont {Stern}}, \bibinfo {author} {\bibfnamefont
  {V.}~\bibnamefont {Falko}}, \bibinfo {author} {\bibfnamefont
  {M.}~\bibnamefont {Goldstein}}, \bibinfo {author} {\bibfnamefont
  {L.}~\bibnamefont {Kronik}}, \bibinfo {author} {\bibfnamefont
  {M.}~\bibnamefont {Urbakh}}, \bibinfo {author} {\bibfnamefont
  {O.}~\bibnamefont {Hod}}, \ and\ \bibinfo {author} {\bibfnamefont {M.~B.}\
  \bibnamefont {Shalom}},\ }\href@noop {} {} (\bibinfo {year} {2023}),\ \Eprint
  {http://arxiv.org/abs/2305.10890} {arXiv:2305.10890 [cond-mat.mtrl-sci]}
  \BibitemShut {NoStop}%
\bibitem [{\citenamefont {Stern}\ \emph {et~al.}(2021)\citenamefont {Stern},
  \citenamefont {Waschitz}, \citenamefont {Cao}, \citenamefont {Nevo},
  \citenamefont {Watanabe}, \citenamefont {Taniguchi}, \citenamefont {Sela},
  \citenamefont {Urbakh}, \citenamefont {Hod},\ and\ \citenamefont
  {Shalom}}]{Vizner2021}%
  \BibitemOpen
  \bibfield  {author} {\bibinfo {author} {\bibfnamefont {M.~V.}\ \bibnamefont
  {Stern}}, \bibinfo {author} {\bibfnamefont {Y.}~\bibnamefont {Waschitz}},
  \bibinfo {author} {\bibfnamefont {W.}~\bibnamefont {Cao}}, \bibinfo {author}
  {\bibfnamefont {I.}~\bibnamefont {Nevo}}, \bibinfo {author} {\bibfnamefont
  {K.}~\bibnamefont {Watanabe}}, \bibinfo {author} {\bibfnamefont
  {T.}~\bibnamefont {Taniguchi}}, \bibinfo {author} {\bibfnamefont
  {E.}~\bibnamefont {Sela}}, \bibinfo {author} {\bibfnamefont {M.}~\bibnamefont
  {Urbakh}}, \bibinfo {author} {\bibfnamefont {O.}~\bibnamefont {Hod}}, \ and\
  \bibinfo {author} {\bibfnamefont {M.~B.}\ \bibnamefont {Shalom}},\ }\href
  {\doibase 10.1126/science.abe8177} {\bibfield  {journal} {\bibinfo  {journal}
  {Science}\ }\textbf {\bibinfo {volume} {372}},\ \bibinfo {pages} {1462}
  (\bibinfo {year} {2021})}\BibitemShut {NoStop}%
\bibitem [{\citenamefont {Yasuda}\ \emph {et~al.}(2021)\citenamefont {Yasuda},
  \citenamefont {Wang}, \citenamefont {Watanabe}, \citenamefont {Taniguchi},\
  and\ \citenamefont {Jarillo-Herrero}}]{Yasuda2021}%
  \BibitemOpen
  \bibfield  {author} {\bibinfo {author} {\bibfnamefont {K.}~\bibnamefont
  {Yasuda}}, \bibinfo {author} {\bibfnamefont {X.}~\bibnamefont {Wang}},
  \bibinfo {author} {\bibfnamefont {K.}~\bibnamefont {Watanabe}}, \bibinfo
  {author} {\bibfnamefont {T.}~\bibnamefont {Taniguchi}}, \ and\ \bibinfo
  {author} {\bibfnamefont {P.}~\bibnamefont {Jarillo-Herrero}},\ }\href
  {\doibase 10.1126/science.abd3230} {\bibfield  {journal} {\bibinfo  {journal}
  {Science}\ }\textbf {\bibinfo {volume} {372}},\ \bibinfo {pages} {1458}
  (\bibinfo {year} {2021})}\BibitemShut {NoStop}%
\bibitem [{\citenamefont {Weston}\ \emph {et~al.}(2022)\citenamefont {Weston},
  \citenamefont {Castanon}, \citenamefont {Enaldiev}, \citenamefont {Ferreira},
  \citenamefont {Bhattacharjee}, \citenamefont {Xu}, \citenamefont
  {Corte-Le{\'o}n}, \citenamefont {Wu}, \citenamefont {Clark}, \citenamefont
  {Summerfield}, \citenamefont {Hashimoto}, \citenamefont {Gao}, \citenamefont
  {Wang}, \citenamefont {Hamer}, \citenamefont {Read}, \citenamefont
  {Fumagalli}, \citenamefont {Kretinin}, \citenamefont {Haigh}, \citenamefont
  {Kazakova}, \citenamefont {Geim}, \citenamefont {Fal'ko},\ and\ \citenamefont
  {Gorbachev}}]{Weston2022}%
  \BibitemOpen
  \bibfield  {author} {\bibinfo {author} {\bibfnamefont {A.}~\bibnamefont
  {Weston}}, \bibinfo {author} {\bibfnamefont {E.}~\bibnamefont {Castanon}},
  \bibinfo {author} {\bibfnamefont {V.}~\bibnamefont {Enaldiev}}, \bibinfo
  {author} {\bibfnamefont {F.}~\bibnamefont {Ferreira}}, \bibinfo {author}
  {\bibfnamefont {S.}~\bibnamefont {Bhattacharjee}}, \bibinfo {author}
  {\bibfnamefont {S.}~\bibnamefont {Xu}}, \bibinfo {author} {\bibfnamefont
  {H.}~\bibnamefont {Corte-Le{\'o}n}}, \bibinfo {author} {\bibfnamefont
  {Z.}~\bibnamefont {Wu}}, \bibinfo {author} {\bibfnamefont {N.}~\bibnamefont
  {Clark}}, \bibinfo {author} {\bibfnamefont {A.}~\bibnamefont {Summerfield}},
  \bibinfo {author} {\bibfnamefont {T.}~\bibnamefont {Hashimoto}}, \bibinfo
  {author} {\bibfnamefont {Y.}~\bibnamefont {Gao}}, \bibinfo {author}
  {\bibfnamefont {W.}~\bibnamefont {Wang}}, \bibinfo {author} {\bibfnamefont
  {M.}~\bibnamefont {Hamer}}, \bibinfo {author} {\bibfnamefont
  {H.}~\bibnamefont {Read}}, \bibinfo {author} {\bibfnamefont {L.}~\bibnamefont
  {Fumagalli}}, \bibinfo {author} {\bibfnamefont {A.}~\bibnamefont {Kretinin}},
  \bibinfo {author} {\bibfnamefont {S.}~\bibnamefont {Haigh}}, \bibinfo
  {author} {\bibfnamefont {O.}~\bibnamefont {Kazakova}}, \bibinfo {author}
  {\bibfnamefont {A.}~\bibnamefont {Geim}}, \bibinfo {author} {\bibfnamefont
  {V.}~\bibnamefont {Fal'ko}}, \ and\ \bibinfo {author} {\bibfnamefont
  {R.}~\bibnamefont {Gorbachev}},\ }\href {\doibase 10.1038/s41565-022-01072-w}
  {\bibfield  {journal} {\bibinfo  {journal} {Nat. Nanotechnol.}\ }\textbf
  {\bibinfo {volume} {17}},\ \bibinfo {pages} {390} (\bibinfo {year}
  {2022})}\BibitemShut {NoStop}%
\bibitem [{\citenamefont {Rogée}\ \emph {et~al.}(2022)\citenamefont {Rogée},
  \citenamefont {Wang}, \citenamefont {Zhang}, \citenamefont {Cai},
  \citenamefont {Wang}, \citenamefont {Chhowalla}, \citenamefont {Ji},\ and\
  \citenamefont {Lau}}]{Rogee2022}%
  \BibitemOpen
  \bibfield  {author} {\bibinfo {author} {\bibfnamefont {L.}~\bibnamefont
  {Rogée}}, \bibinfo {author} {\bibfnamefont {L.}~\bibnamefont {Wang}},
  \bibinfo {author} {\bibfnamefont {Y.}~\bibnamefont {Zhang}}, \bibinfo
  {author} {\bibfnamefont {S.}~\bibnamefont {Cai}}, \bibinfo {author}
  {\bibfnamefont {P.}~\bibnamefont {Wang}}, \bibinfo {author} {\bibfnamefont
  {M.}~\bibnamefont {Chhowalla}}, \bibinfo {author} {\bibfnamefont
  {W.}~\bibnamefont {Ji}}, \ and\ \bibinfo {author} {\bibfnamefont {S.~P.}\
  \bibnamefont {Lau}},\ }\href {\doibase 10.1126/science.abm5734} {\bibfield
  {journal} {\bibinfo  {journal} {Science}\ }\textbf {\bibinfo {volume}
  {376}},\ \bibinfo {pages} {973} (\bibinfo {year} {2022})}\BibitemShut
  {NoStop}%
\bibitem [{\citenamefont {Woods}\ \emph {et~al.}(2021)\citenamefont {Woods},
  \citenamefont {Ares}, \citenamefont {Nevison-Andrews}, \citenamefont
  {Holwill}, \citenamefont {Fabregas}, \citenamefont {Guinea}, \citenamefont
  {Geim}, \citenamefont {Novoselov}, \citenamefont {Walet},\ and\ \citenamefont
  {Fumagalli}}]{Woods2021}%
  \BibitemOpen
  \bibfield  {author} {\bibinfo {author} {\bibfnamefont {C.~R.}\ \bibnamefont
  {Woods}}, \bibinfo {author} {\bibfnamefont {P.}~\bibnamefont {Ares}},
  \bibinfo {author} {\bibfnamefont {H.}~\bibnamefont {Nevison-Andrews}},
  \bibinfo {author} {\bibfnamefont {M.~J.}\ \bibnamefont {Holwill}}, \bibinfo
  {author} {\bibfnamefont {R.}~\bibnamefont {Fabregas}}, \bibinfo {author}
  {\bibfnamefont {F.}~\bibnamefont {Guinea}}, \bibinfo {author} {\bibfnamefont
  {A.~K.}\ \bibnamefont {Geim}}, \bibinfo {author} {\bibfnamefont {K.~S.}\
  \bibnamefont {Novoselov}}, \bibinfo {author} {\bibfnamefont {N.~R.}\
  \bibnamefont {Walet}}, \ and\ \bibinfo {author} {\bibfnamefont
  {L.}~\bibnamefont {Fumagalli}},\ }\href {\doibase 10.1038/s41467-020-20667-2}
  {\bibfield  {journal} {\bibinfo  {journal} {Nat. Comm.}\ }\textbf {\bibinfo
  {volume} {12}},\ \bibinfo {pages} {347} (\bibinfo {year} {2021})}\BibitemShut
  {NoStop}%
\bibitem [{\citenamefont {Garcia-Ruiz}\ \emph {et~al.}(2023)\citenamefont
  {Garcia-Ruiz}, \citenamefont {Enaldiev}, \citenamefont {McEllistrim},\ and\
  \citenamefont {Fal'ko}}]{garciaruiz2023}%
  \BibitemOpen
  \bibfield  {author} {\bibinfo {author} {\bibfnamefont {A.}~\bibnamefont
  {Garcia-Ruiz}}, \bibinfo {author} {\bibfnamefont {V.}~\bibnamefont
  {Enaldiev}}, \bibinfo {author} {\bibfnamefont {A.}~\bibnamefont
  {McEllistrim}}, \ and\ \bibinfo {author} {\bibfnamefont {V.~I.}\ \bibnamefont
  {Fal'ko}},\ }\href@noop {} {} (\bibinfo {year} {2023}),\ \Eprint
  {http://arxiv.org/abs/2305.10896} {arXiv:2305.10896 [cond-mat.mtrl-sci]}
  \BibitemShut {NoStop}%
\bibitem [{\citenamefont {Carr}\ \emph {et~al.}(2018)\citenamefont {Carr},
  \citenamefont {Massatt}, \citenamefont {Torrisi}, \citenamefont {Cazeaux},
  \citenamefont {Luskin},\ and\ \citenamefont {Kaxiras}}]{Carr2018}%
  \BibitemOpen
  \bibfield  {author} {\bibinfo {author} {\bibfnamefont {S.}~\bibnamefont
  {Carr}}, \bibinfo {author} {\bibfnamefont {D.}~\bibnamefont {Massatt}},
  \bibinfo {author} {\bibfnamefont {S.~B.}\ \bibnamefont {Torrisi}}, \bibinfo
  {author} {\bibfnamefont {P.}~\bibnamefont {Cazeaux}}, \bibinfo {author}
  {\bibfnamefont {M.}~\bibnamefont {Luskin}}, \ and\ \bibinfo {author}
  {\bibfnamefont {E.}~\bibnamefont {Kaxiras}},\ }\href {\doibase
  10.1103/PhysRevB.98.224102} {\bibfield  {journal} {\bibinfo  {journal} {Phys.
  Rev. B}\ }\textbf {\bibinfo {volume} {98}},\ \bibinfo {pages} {224102}
  (\bibinfo {year} {2018})}\BibitemShut {NoStop}%
\bibitem [{\citenamefont {Gargiulo}\ and\ \citenamefont
  {Yazyev}(2017)}]{Gargiulo2018}%
  \BibitemOpen
  \bibfield  {author} {\bibinfo {author} {\bibfnamefont {F.}~\bibnamefont
  {Gargiulo}}\ and\ \bibinfo {author} {\bibfnamefont {O.~V.}\ \bibnamefont
  {Yazyev}},\ }\href {\doibase 10.1088/2053-1583/aa9640} {\bibfield  {journal}
  {\bibinfo  {journal} {2D Mater.}\ }\textbf {\bibinfo {volume} {5}},\ \bibinfo
  {pages} {015019} (\bibinfo {year} {2017})}\BibitemShut {NoStop}%
\bibitem [{\citenamefont {Rosenberger}\ \emph {et~al.}(2020)\citenamefont
  {Rosenberger}, \citenamefont {Chuang}, \citenamefont {Phillips},
  \citenamefont {Oleshko}, \citenamefont {McCreary}, \citenamefont {Sivaram},
  \citenamefont {Hellberg},\ and\ \citenamefont {Jonker}}]{Rosenberger2020}%
  \BibitemOpen
  \bibfield  {author} {\bibinfo {author} {\bibfnamefont {M.}~\bibnamefont
  {Rosenberger}}, \bibinfo {author} {\bibfnamefont {H.}~\bibnamefont {Chuang}},
  \bibinfo {author} {\bibfnamefont {M.}~\bibnamefont {Phillips}}, \bibinfo
  {author} {\bibfnamefont {V.}~\bibnamefont {Oleshko}}, \bibinfo {author}
  {\bibfnamefont {K.}~\bibnamefont {McCreary}}, \bibinfo {author}
  {\bibfnamefont {S.}~\bibnamefont {Sivaram}}, \bibinfo {author} {\bibfnamefont
  {C.}~\bibnamefont {Hellberg}}, \ and\ \bibinfo {author} {\bibfnamefont
  {B.}~\bibnamefont {Jonker}},\ }\href {\doibase 10.1021/acsnano.0c00088}
  {\bibfield  {journal} {\bibinfo  {journal} {ACS Nano}\ }\textbf {\bibinfo
  {volume} {14}},\ \bibinfo {pages} {4550} (\bibinfo {year}
  {2020})}\BibitemShut {NoStop}%
\bibitem [{\citenamefont {Weston}\ \emph {et~al.}(2020)\citenamefont {Weston},
  \citenamefont {Zou}, \citenamefont {Enaldiev}, \citenamefont {Summerfield},
  \citenamefont {Clark}, \citenamefont {Z{\'o}lyomi}, \citenamefont {Graham},
  \citenamefont {Yelgel}, \citenamefont {Magorrian}, \citenamefont {Zhou},
  \citenamefont {Zultak}, \citenamefont {Hopkinson}, \citenamefont {Barinov},
  \citenamefont {Bointon}, \citenamefont {Kretinin}, \citenamefont {Wilson},
  \citenamefont {Beton}, \citenamefont {Fal'ko}, \citenamefont {Haigh},\ and\
  \citenamefont {Gorbachev}}]{Weston2020}%
  \BibitemOpen
  \bibfield  {author} {\bibinfo {author} {\bibfnamefont {A.}~\bibnamefont
  {Weston}}, \bibinfo {author} {\bibfnamefont {Y.}~\bibnamefont {Zou}},
  \bibinfo {author} {\bibfnamefont {V.}~\bibnamefont {Enaldiev}}, \bibinfo
  {author} {\bibfnamefont {A.}~\bibnamefont {Summerfield}}, \bibinfo {author}
  {\bibfnamefont {N.}~\bibnamefont {Clark}}, \bibinfo {author} {\bibfnamefont
  {V.}~\bibnamefont {Z{\'o}lyomi}}, \bibinfo {author} {\bibfnamefont
  {A.}~\bibnamefont {Graham}}, \bibinfo {author} {\bibfnamefont
  {C.}~\bibnamefont {Yelgel}}, \bibinfo {author} {\bibfnamefont
  {S.}~\bibnamefont {Magorrian}}, \bibinfo {author} {\bibfnamefont
  {M.}~\bibnamefont {Zhou}}, \bibinfo {author} {\bibfnamefont {J.}~\bibnamefont
  {Zultak}}, \bibinfo {author} {\bibfnamefont {D.}~\bibnamefont {Hopkinson}},
  \bibinfo {author} {\bibfnamefont {A.}~\bibnamefont {Barinov}}, \bibinfo
  {author} {\bibfnamefont {T.}~\bibnamefont {Bointon}}, \bibinfo {author}
  {\bibfnamefont {A.}~\bibnamefont {Kretinin}}, \bibinfo {author}
  {\bibfnamefont {N.}~\bibnamefont {Wilson}}, \bibinfo {author} {\bibfnamefont
  {P.}~\bibnamefont {Beton}}, \bibinfo {author} {\bibfnamefont
  {V.}~\bibnamefont {Fal'ko}}, \bibinfo {author} {\bibfnamefont
  {S.}~\bibnamefont {Haigh}}, \ and\ \bibinfo {author} {\bibfnamefont
  {R.}~\bibnamefont {Gorbachev}},\ }\href {\doibase 10.1038/s41565-020-0682-9}
  {\bibfield  {journal} {\bibinfo  {journal} {Nat. Nanotechnol.}\ }\textbf
  {\bibinfo {volume} {15}},\ \bibinfo {pages} {592} (\bibinfo {year}
  {2020})}\BibitemShut {NoStop}%
\bibitem [{\citenamefont {Lau}\ \emph {et~al.}(2022)\citenamefont {Lau},
  \citenamefont {Bockrath}, \citenamefont {Mak},\ and\ \citenamefont
  {Zhang}}]{Lau2022}%
  \BibitemOpen
  \bibfield  {author} {\bibinfo {author} {\bibfnamefont {C.}~\bibnamefont
  {Lau}}, \bibinfo {author} {\bibfnamefont {M.}~\bibnamefont {Bockrath}},
  \bibinfo {author} {\bibfnamefont {K.~F.}\ \bibnamefont {Mak}}, \ and\
  \bibinfo {author} {\bibfnamefont {F.}~\bibnamefont {Zhang}},\ }\href
  {\doibase 10.1038/s41586-021-04173-z} {\bibfield  {journal} {\bibinfo
  {journal} {Nature}\ }\textbf {\bibinfo {volume} {602}},\ \bibinfo {pages}
  {41} (\bibinfo {year} {2022})}\BibitemShut {NoStop}%
\bibitem [{\citenamefont {Miao}\ \emph {et~al.}(2022)\citenamefont {Miao},
  \citenamefont {Ding}, \citenamefont {Wang}, \citenamefont {Shi},
  \citenamefont {Ye}, \citenamefont {Li}, \citenamefont {Yao}, \citenamefont
  {Dong},\ and\ \citenamefont {Zhang}}]{Miao2022}%
  \BibitemOpen
  \bibfield  {author} {\bibinfo {author} {\bibfnamefont {L.}~\bibnamefont
  {Miao}}, \bibinfo {author} {\bibfnamefont {N.}~\bibnamefont {Ding}}, \bibinfo
  {author} {\bibfnamefont {N.}~\bibnamefont {Wang}}, \bibinfo {author}
  {\bibfnamefont {C.}~\bibnamefont {Shi}}, \bibinfo {author} {\bibfnamefont
  {H.-Y.}\ \bibnamefont {Ye}}, \bibinfo {author} {\bibfnamefont
  {L.}~\bibnamefont {Li}}, \bibinfo {author} {\bibfnamefont {Y.}~\bibnamefont
  {Yao}}, \bibinfo {author} {\bibfnamefont {S.}~\bibnamefont {Dong}}, \ and\
  \bibinfo {author} {\bibfnamefont {Y.}~\bibnamefont {Zhang}},\ }\href
  {\doibase 10.1038/s41563-022-01322-1} {\bibfield  {journal} {\bibinfo
  {journal} {Nat. Mater.}\ }\textbf {\bibinfo {volume} {21}},\ \bibinfo {pages}
  {1158} (\bibinfo {year} {2022})}\BibitemShut {NoStop}%
\bibitem [{\citenamefont {Suzuki}\ \emph {et~al.}(2014)\citenamefont {Suzuki},
  \citenamefont {Sakano}, \citenamefont {Zhang}, \citenamefont {Akashi},
  \citenamefont {Morikawa}, \citenamefont {Harasawa}, \citenamefont {Yaji},
  \citenamefont {Kuroda}, \citenamefont {Miyamoto}, \citenamefont {Okuda},
  \citenamefont {Ishizaka}, \citenamefont {Arita},\ and\ \citenamefont
  {Iwasa}}]{Suzuki2014}%
  \BibitemOpen
  \bibfield  {author} {\bibinfo {author} {\bibfnamefont {R.}~\bibnamefont
  {Suzuki}}, \bibinfo {author} {\bibfnamefont {M.}~\bibnamefont {Sakano}},
  \bibinfo {author} {\bibfnamefont {Y.}~\bibnamefont {Zhang}}, \bibinfo
  {author} {\bibfnamefont {R.}~\bibnamefont {Akashi}}, \bibinfo {author}
  {\bibfnamefont {D.}~\bibnamefont {Morikawa}}, \bibinfo {author}
  {\bibfnamefont {A.}~\bibnamefont {Harasawa}}, \bibinfo {author}
  {\bibfnamefont {K.}~\bibnamefont {Yaji}}, \bibinfo {author} {\bibfnamefont
  {K.}~\bibnamefont {Kuroda}}, \bibinfo {author} {\bibfnamefont
  {K.}~\bibnamefont {Miyamoto}}, \bibinfo {author} {\bibfnamefont
  {T.}~\bibnamefont {Okuda}}, \bibinfo {author} {\bibfnamefont
  {K.}~\bibnamefont {Ishizaka}}, \bibinfo {author} {\bibfnamefont
  {R.}~\bibnamefont {Arita}}, \ and\ \bibinfo {author} {\bibfnamefont
  {Y.}~\bibnamefont {Iwasa}},\ }\href {\doibase 10.1038/nnano.2014.148}
  {\bibfield  {journal} {\bibinfo  {journal} {Nat. Nanotechnol.}\ }\textbf
  {\bibinfo {volume} {9}},\ \bibinfo {pages} {611} (\bibinfo {year}
  {2014})}\BibitemShut {NoStop}%
\bibitem [{\citenamefont {Mortelmans}\ \emph {et~al.}(2021)\citenamefont
  {Mortelmans}, \citenamefont {{De Gendt}}, \citenamefont {Heyns},\ and\
  \citenamefont {Merckling}}]{Mortelmans2021}%
  \BibitemOpen
  \bibfield  {author} {\bibinfo {author} {\bibfnamefont {W.}~\bibnamefont
  {Mortelmans}}, \bibinfo {author} {\bibfnamefont {S.}~\bibnamefont {{De
  Gendt}}}, \bibinfo {author} {\bibfnamefont {M.}~\bibnamefont {Heyns}}, \ and\
  \bibinfo {author} {\bibfnamefont {C.}~\bibnamefont {Merckling}},\ }\href
  {\doibase 10.1016/j.apmt.2021.100975} {\bibfield  {journal} {\bibinfo
  {journal} {Appl. Mater. Today}\ }\textbf {\bibinfo {volume} {22}},\ \bibinfo
  {pages} {100975} (\bibinfo {year} {2021})}\BibitemShut {NoStop}%
\bibitem [{\citenamefont {Wiegers}(1996)}]{Wiegers1996}%
  \BibitemOpen
  \bibfield  {author} {\bibinfo {author} {\bibfnamefont {G.}~\bibnamefont
  {Wiegers}},\ }\href {\doibase 10.1016/0079-6786(95)00007-0} {\bibfield
  {journal} {\bibinfo  {journal} {Prog. Solid. State Ch.}\ }\textbf {\bibinfo
  {volume} {24}},\ \bibinfo {pages} {1} (\bibinfo {year} {1996})}\BibitemShut
  {NoStop}%
\bibitem [{\citenamefont {Pet\v{r}\'{i}\v{c}ek}\ \emph
  {et~al.}(1991)\citenamefont {Pet\v{r}\'{i}\v{c}ek}, \citenamefont {Mal\'{y}},
  \citenamefont {Coppens}, \citenamefont {Bu}, \citenamefont
  {C\'{i}sa\v{r}ov\'{a}},\ and\ \citenamefont {Frost-Jensen}}]{Vaclav1991}%
  \BibitemOpen
  \bibfield  {author} {\bibinfo {author} {\bibfnamefont {V.}~\bibnamefont
  {Pet\v{r}\'{i}\v{c}ek}}, \bibinfo {author} {\bibfnamefont {K.}~\bibnamefont
  {Mal\'{y}}}, \bibinfo {author} {\bibfnamefont {P.}~\bibnamefont {Coppens}},
  \bibinfo {author} {\bibfnamefont {X.}~\bibnamefont {Bu}}, \bibinfo {author}
  {\bibfnamefont {I.}~\bibnamefont {C\'{i}sa\v{r}ov\'{a}}}, \ and\ \bibinfo
  {author} {\bibfnamefont {A.}~\bibnamefont {Frost-Jensen}},\ }\href {\doibase
  10.1107/S0108767392009437} {\bibfield  {journal} {\bibinfo  {journal} {Acta
  Crystallogr. Sect. A}\ }\textbf {\bibinfo {volume} {47}},\ \bibinfo {pages}
  {210} (\bibinfo {year} {1991})}\BibitemShut {NoStop}%
\bibitem [{\citenamefont {C\'{i}sa\v{r}ov\'{a}}\ \emph
  {et~al.}(1993)\citenamefont {C\'{i}sa\v{r}ov\'{a}}, \citenamefont {Mal\'{y}},
  \citenamefont {Pet\v{r}\'{i}\v{c}ek},\ and\ \citenamefont
  {Coppens}}]{Cisarova1993}%
  \BibitemOpen
  \bibfield  {author} {\bibinfo {author} {\bibfnamefont {I.}~\bibnamefont
  {C\'{i}sa\v{r}ov\'{a}}}, \bibinfo {author} {\bibfnamefont {K.}~\bibnamefont
  {Mal\'{y}}}, \bibinfo {author} {\bibfnamefont {V.}~\bibnamefont
  {Pet\v{r}\'{i}\v{c}ek}}, \ and\ \bibinfo {author} {\bibfnamefont
  {P.}~\bibnamefont {Coppens}},\ }\href {\doibase 10.1107/S0108767392009437}
  {\bibfield  {journal} {\bibinfo  {journal} {Acta Crystallogr. Sect. A}\
  }\textbf {\bibinfo {volume} {49}},\ \bibinfo {pages} {336} (\bibinfo {year}
  {1993})}\BibitemShut {NoStop}%
\bibitem [{\citenamefont {Oxford~UK}(2022)}]{crysalis}%
  \BibitemOpen
  \bibfield  {author} {\bibinfo {author} {\bibfnamefont {R.~C.}\ \bibnamefont
  {Oxford~UK}},\ }\href@noop {} {\enquote {\bibinfo {title} {{Rigaku Oxford
  Diffraction, CrysAlisPro Software system}},}\ } (\bibinfo {year}
  {2022})\BibitemShut {NoStop}%
\bibitem [{\citenamefont {Gotoh}\ \emph {et~al.}(1990)\citenamefont {Gotoh},
  \citenamefont {Goto}, \citenamefont {Kawaguchi}, \citenamefont {Oosawa},\
  and\ \citenamefont {Onoda}}]{Gotoh1990}%
  \BibitemOpen
  \bibfield  {author} {\bibinfo {author} {\bibfnamefont {Y.}~\bibnamefont
  {Gotoh}}, \bibinfo {author} {\bibfnamefont {M.}~\bibnamefont {Goto}},
  \bibinfo {author} {\bibfnamefont {K.}~\bibnamefont {Kawaguchi}}, \bibinfo
  {author} {\bibfnamefont {Y.}~\bibnamefont {Oosawa}}, \ and\ \bibinfo {author}
  {\bibfnamefont {M.}~\bibnamefont {Onoda}},\ }\href {\doibase
  10.1016/0025-5408(90)90101-7} {\bibfield  {journal} {\bibinfo  {journal}
  {Mater. Res. Bull.}\ }\textbf {\bibinfo {volume} {25}},\ \bibinfo {pages}
  {307} (\bibinfo {year} {1990})}\BibitemShut {NoStop}%
\bibitem [{sup(2023)}]{suppl}%
  \BibitemOpen
  \href@noop {} {\enquote {\bibinfo {title} {See supplemental material at xxx
  for experimental details, {XPS} core level spectroscopy, {TEM} and the
  reciprocal space map of {(PbS)$_{1.18}$VS$_2$} , which includes refs.
  [40-55].}}\ } (\bibinfo {year} {2023})\BibitemShut {NoStop}%
\bibitem [{\citenamefont {Mitsuko}\ \emph {et~al.}(1990)\citenamefont
  {Mitsuko}, \citenamefont {Katsuo}, \citenamefont {Yoshito},\ and\
  \citenamefont {Yoshinao}}]{Onoda1990}%
  \BibitemOpen
  \bibfield  {author} {\bibinfo {author} {\bibfnamefont {O.}~\bibnamefont
  {Mitsuko}}, \bibinfo {author} {\bibfnamefont {K.}~\bibnamefont {Katsuo}},
  \bibinfo {author} {\bibfnamefont {G.}~\bibnamefont {Yoshito}}, \ and\
  \bibinfo {author} {\bibfnamefont {O.}~\bibnamefont {Yoshinao}},\ }\href
  {\doibase 10.1107/S0108768190003950} {\bibfield  {journal} {\bibinfo
  {journal} {Acta Crystallogr. Sect. B}\ }\textbf {\bibinfo {volume} {46}},\
  \bibinfo {pages} {487} (\bibinfo {year} {1990})}\BibitemShut {NoStop}%
\bibitem [{cry(2023)}]{crystal_structure}%
  \BibitemOpen
  \href@noop {} {\enquote {\bibinfo {title} {A detailed description of the
  crystal structure is under preparation and will be published elsewhere.}}\ }
  (\bibinfo {year} {2023})\BibitemShut {NoStop}%
\bibitem [{\citenamefont {Andersen}\ \emph {et~al.}(2021)\citenamefont
  {Andersen}, \citenamefont {Scuri}, \citenamefont {Sushko}, \citenamefont
  {De~Greve}, \citenamefont {Sung}, \citenamefont {Zhou}, \citenamefont {Wild},
  \citenamefont {Gelly}, \citenamefont {Heo}, \citenamefont {B{\'e}rub{\'e}},
  \citenamefont {Joe}, \citenamefont {Jauregui}, \citenamefont {Watanabe},
  \citenamefont {Taniguchi}, \citenamefont {Kim}, \citenamefont {Park},\ and\
  \citenamefont {Lukin}}]{Andersen2021}%
  \BibitemOpen
  \bibfield  {author} {\bibinfo {author} {\bibfnamefont {T.~I.}\ \bibnamefont
  {Andersen}}, \bibinfo {author} {\bibfnamefont {G.}~\bibnamefont {Scuri}},
  \bibinfo {author} {\bibfnamefont {A.}~\bibnamefont {Sushko}}, \bibinfo
  {author} {\bibfnamefont {K.}~\bibnamefont {De~Greve}}, \bibinfo {author}
  {\bibfnamefont {J.}~\bibnamefont {Sung}}, \bibinfo {author} {\bibfnamefont
  {Y.}~\bibnamefont {Zhou}}, \bibinfo {author} {\bibfnamefont {D.~S.}\
  \bibnamefont {Wild}}, \bibinfo {author} {\bibfnamefont {R.~J.}\ \bibnamefont
  {Gelly}}, \bibinfo {author} {\bibfnamefont {H.}~\bibnamefont {Heo}}, \bibinfo
  {author} {\bibfnamefont {D.}~\bibnamefont {B{\'e}rub{\'e}}}, \bibinfo
  {author} {\bibfnamefont {A.~Y.}\ \bibnamefont {Joe}}, \bibinfo {author}
  {\bibfnamefont {L.~A.}\ \bibnamefont {Jauregui}}, \bibinfo {author}
  {\bibfnamefont {K.}~\bibnamefont {Watanabe}}, \bibinfo {author}
  {\bibfnamefont {T.}~\bibnamefont {Taniguchi}}, \bibinfo {author}
  {\bibfnamefont {P.}~\bibnamefont {Kim}}, \bibinfo {author} {\bibfnamefont
  {H.}~\bibnamefont {Park}}, \ and\ \bibinfo {author} {\bibfnamefont {M.~D.}\
  \bibnamefont {Lukin}},\ }\href {\doibase 10.1038/s41563-020-00873-5}
  {\bibfield  {journal} {\bibinfo  {journal} {Nat. Mat.}\ }\textbf {\bibinfo
  {volume} {20}},\ \bibinfo {pages} {480} (\bibinfo {year} {2021})}\BibitemShut
  {NoStop}%
\bibitem [{\citenamefont {Bihan}\ and\ \citenamefont
  {Maussion}(1972)}]{LEBIHAN1972}%
  \BibitemOpen
  \bibfield  {author} {\bibinfo {author} {\bibfnamefont {R.~L.}\ \bibnamefont
  {Bihan}}\ and\ \bibinfo {author} {\bibfnamefont {M.}~\bibnamefont
  {Maussion}},\ }\href {\doibase 10.1051/jphyscol:1972274} {\bibfield
  {journal} {\bibinfo  {journal} {J. Phys. Colloques}\ }\textbf {\bibinfo
  {volume} {33}},\ \bibinfo {pages} {C2} (\bibinfo {year} {1972})}\BibitemShut
  {NoStop}%
\bibitem [{\citenamefont {Hunnestad}\ \emph {et~al.}(2020)\citenamefont
  {Hunnestad}, \citenamefont {Roede}, \citenamefont {van Helvoort},\ and\
  \citenamefont {Meier}}]{Hunnestad2020}%
  \BibitemOpen
  \bibfield  {author} {\bibinfo {author} {\bibfnamefont {K.}~\bibnamefont
  {Hunnestad}}, \bibinfo {author} {\bibfnamefont {E.}~\bibnamefont {Roede}},
  \bibinfo {author} {\bibfnamefont {A.}~\bibnamefont {van Helvoort}}, \ and\
  \bibinfo {author} {\bibfnamefont {D.}~\bibnamefont {Meier}},\ }\href
  {\doibase 10.1063/5.0029284} {\bibfield  {journal} {\bibinfo  {journal} {J.
  Appl. Phys.}\ }\textbf {\bibinfo {volume} {128}},\ \bibinfo {pages} {191102}
  (\bibinfo {year} {2020})}\BibitemShut {NoStop}%
\bibitem [{\citenamefont {Moore}\ \emph {et~al.}(2021)\citenamefont {Moore},
  \citenamefont {Ciccarino}, \citenamefont {Halbertal}, \citenamefont
  {McGilly}, \citenamefont {Finney}, \citenamefont {Yao}, \citenamefont {Shao},
  \citenamefont {Ni}, \citenamefont {Sternbach}, \citenamefont {Telford},
  \citenamefont {Kim}, \citenamefont {Rossi}, \citenamefont {Watanabe},
  \citenamefont {Taniguchi}, \citenamefont {Pasupathy}, \citenamefont {Dean},
  \citenamefont {Hone}, \citenamefont {Schuck}, \citenamefont {Narang},\ and\
  \citenamefont {Basov}}]{Moore2021}%
  \BibitemOpen
  \bibfield  {author} {\bibinfo {author} {\bibfnamefont {S.~L.}\ \bibnamefont
  {Moore}}, \bibinfo {author} {\bibfnamefont {C.}~\bibnamefont {Ciccarino}},
  \bibinfo {author} {\bibfnamefont {D.}~\bibnamefont {Halbertal}}, \bibinfo
  {author} {\bibfnamefont {L.}~\bibnamefont {McGilly}}, \bibinfo {author}
  {\bibfnamefont {N.}~\bibnamefont {Finney}}, \bibinfo {author} {\bibfnamefont
  {K.}~\bibnamefont {Yao}}, \bibinfo {author} {\bibfnamefont {Y.}~\bibnamefont
  {Shao}}, \bibinfo {author} {\bibfnamefont {G.}~\bibnamefont {Ni}}, \bibinfo
  {author} {\bibfnamefont {A.}~\bibnamefont {Sternbach}}, \bibinfo {author}
  {\bibfnamefont {E.}~\bibnamefont {Telford}}, \bibinfo {author} {\bibfnamefont
  {B.}~\bibnamefont {Kim}}, \bibinfo {author} {\bibfnamefont {S.}~\bibnamefont
  {Rossi}}, \bibinfo {author} {\bibfnamefont {K.}~\bibnamefont {Watanabe}},
  \bibinfo {author} {\bibfnamefont {T.}~\bibnamefont {Taniguchi}}, \bibinfo
  {author} {\bibfnamefont {A.}~\bibnamefont {Pasupathy}}, \bibinfo {author}
  {\bibfnamefont {C.}~\bibnamefont {Dean}}, \bibinfo {author} {\bibfnamefont
  {J.}~\bibnamefont {Hone}}, \bibinfo {author} {\bibfnamefont {P.}~\bibnamefont
  {Schuck}}, \bibinfo {author} {\bibfnamefont {P.}~\bibnamefont {Narang}}, \
  and\ \bibinfo {author} {\bibfnamefont {D.}~\bibnamefont {Basov}},\ }\href
  {\doibase 10.1038/s41467-021-26072-7} {\bibfield  {journal} {\bibinfo
  {journal} {Nat. Commun.}\ }\textbf {\bibinfo {volume} {12}},\ \bibinfo
  {pages} {5741} (\bibinfo {year} {2021})}\BibitemShut {NoStop}%
\bibitem [{\citenamefont {Chiodini}\ \emph {et~al.}(2022)\citenamefont
  {Chiodini}, \citenamefont {Kerfoot}, \citenamefont {Venturi}, \citenamefont
  {Mignuzzi}, \citenamefont {Alexeev}, \citenamefont {Rosa}, \citenamefont
  {Tongay}, \citenamefont {Taniguchi}, \citenamefont {Watanabe}, \citenamefont
  {Ferrari},\ and\ \citenamefont {Ambrosio}}]{Chiodini2022}%
  \BibitemOpen
  \bibfield  {author} {\bibinfo {author} {\bibfnamefont {S.}~\bibnamefont
  {Chiodini}}, \bibinfo {author} {\bibfnamefont {J.}~\bibnamefont {Kerfoot}},
  \bibinfo {author} {\bibfnamefont {G.}~\bibnamefont {Venturi}}, \bibinfo
  {author} {\bibfnamefont {S.}~\bibnamefont {Mignuzzi}}, \bibinfo {author}
  {\bibfnamefont {E.}~\bibnamefont {Alexeev}}, \bibinfo {author} {\bibfnamefont
  {B.~T.}\ \bibnamefont {Rosa}}, \bibinfo {author} {\bibfnamefont
  {S.}~\bibnamefont {Tongay}}, \bibinfo {author} {\bibfnamefont
  {T.}~\bibnamefont {Taniguchi}}, \bibinfo {author} {\bibfnamefont
  {K.}~\bibnamefont {Watanabe}}, \bibinfo {author} {\bibfnamefont
  {A.}~\bibnamefont {Ferrari}}, \ and\ \bibinfo {author} {\bibfnamefont
  {A.}~\bibnamefont {Ambrosio}},\ }\href {\doibase 10.1021/acsnano.1c11107}
  {\bibfield  {journal} {\bibinfo  {journal} {ACS Nano}\ }\textbf {\bibinfo
  {volume} {16}},\ \bibinfo {pages} {7589} (\bibinfo {year}
  {2022})}\BibitemShut {NoStop}%
\bibitem [{\citenamefont {Bihan}(1970)}]{Bihan1970}%
  \BibitemOpen
  \bibfield  {author} {\bibinfo {author} {\bibfnamefont {R.~L.}\ \bibnamefont
  {Bihan}},\ }\href@noop {} {\bibfield  {journal} {\bibinfo  {journal} {C. R.
  Acad. Sci. Ser. B}\ }\textbf {\bibinfo {volume} {270}},\ \bibinfo {pages}
  {741} (\bibinfo {year} {1970})}\BibitemShut {NoStop}%
\bibitem [{\citenamefont {Bihan}(1972)}]{Bihan1972}%
  \BibitemOpen
  \bibfield  {author} {\bibinfo {author} {\bibfnamefont {R.~L.}\ \bibnamefont
  {Bihan}},\ }\href@noop {} {\bibfield  {journal} {\bibinfo  {journal} {C. R.
  Acad. Sci. Ser. B}\ }\textbf {\bibinfo {volume} {275}},\ \bibinfo {pages}
  {29} (\bibinfo {year} {1972})}\BibitemShut {NoStop}%
\bibitem [{\citenamefont {Morlon}\ \emph {et~al.}(1970)\citenamefont {Morlon},
  \citenamefont {Coquet},\ and\ \citenamefont {Devin}}]{Morlon1970}%
  \BibitemOpen
  \bibfield  {author} {\bibinfo {author} {\bibfnamefont {B.}~\bibnamefont
  {Morlon}}, \bibinfo {author} {\bibfnamefont {E.}~\bibnamefont {Coquet}}, \
  and\ \bibinfo {author} {\bibfnamefont {A.}~\bibnamefont {Devin}},\
  }\href@noop {} {\bibfield  {journal} {\bibinfo  {journal} {C. R. Acad. Sci.
  Ser. B}\ }\textbf {\bibinfo {volume} {270}},\ \bibinfo {pages} {283}
  (\bibinfo {year} {1970})}\BibitemShut {NoStop}%
\bibitem [{\citenamefont {Sourisseau}\ \emph {et~al.}(1995)\citenamefont
  {Sourisseau}, \citenamefont {Cavagnat}, \citenamefont {Fouassier},
  \citenamefont {Tirado},\ and\ \citenamefont {Morales}}]{Sourisseau1995}%
  \BibitemOpen
  \bibfield  {author} {\bibinfo {author} {\bibfnamefont {C.}~\bibnamefont
  {Sourisseau}}, \bibinfo {author} {\bibfnamefont {R.}~\bibnamefont
  {Cavagnat}}, \bibinfo {author} {\bibfnamefont {M.}~\bibnamefont {Fouassier}},
  \bibinfo {author} {\bibfnamefont {J.~L.}\ \bibnamefont {Tirado}}, \ and\
  \bibinfo {author} {\bibfnamefont {J.}~\bibnamefont {Morales}},\ }\href
  {\doibase 10.1002/jrs.1250260814} {\bibfield  {journal} {\bibinfo  {journal}
  {J. Raman Spectrosc.}\ }\textbf {\bibinfo {volume} {26}},\ \bibinfo {pages}
  {675} (\bibinfo {year} {1995})}\BibitemShut {NoStop}%
\bibitem [{\citenamefont {Ng}\ and\ \citenamefont {McQueen}(2022)}]{Ng2022}%
  \BibitemOpen
  \bibfield  {author} {\bibinfo {author} {\bibfnamefont {N.}~\bibnamefont
  {Ng}}\ and\ \bibinfo {author} {\bibfnamefont {T.}~\bibnamefont {McQueen}},\
  }\href {\doibase 10.1063/5.0101429} {\bibfield  {journal} {\bibinfo
  {journal} {APL Mater.}\ }\textbf {\bibinfo {volume} {10}},\ \bibinfo {pages}
  {100901} (\bibinfo {year} {2022})}\BibitemShut {NoStop}%
\bibitem [{\citenamefont {Cosma}\ \emph {et~al.}(2014)\citenamefont {Cosma},
  \citenamefont {Wallbank}, \citenamefont {Cheianov},\ and\ \citenamefont
  {Fal{'}ko}}]{Cosma2014}%
  \BibitemOpen
  \bibfield  {author} {\bibinfo {author} {\bibfnamefont {D.~A.}\ \bibnamefont
  {Cosma}}, \bibinfo {author} {\bibfnamefont {J.}~\bibnamefont {Wallbank}},
  \bibinfo {author} {\bibfnamefont {V.}~\bibnamefont {Cheianov}}, \ and\
  \bibinfo {author} {\bibfnamefont {V.}~\bibnamefont {Fal{'}ko}},\ }\href
  {\doibase 10.1039/C4FD00146J} {\bibfield  {journal} {\bibinfo  {journal}
  {Faraday Discuss.}\ }\textbf {\bibinfo {volume} {173}},\ \bibinfo {pages}
  {137} (\bibinfo {year} {2014})}\BibitemShut {NoStop}%
\bibitem [{\citenamefont {Ino}\ \emph {et~al.}(2004)\citenamefont {Ino},
  \citenamefont {Okane}, \citenamefont {Fujimori}, \citenamefont {Fujimori},
  \citenamefont {Mizokawa}, \citenamefont {Yasui}, \citenamefont {Nishikawa},\
  and\ \citenamefont {Sato}}]{Ino2004}%
  \BibitemOpen
  \bibfield  {author} {\bibinfo {author} {\bibfnamefont {A.}~\bibnamefont
  {Ino}}, \bibinfo {author} {\bibfnamefont {T.}~\bibnamefont {Okane}}, \bibinfo
  {author} {\bibfnamefont {S.-I.}\ \bibnamefont {Fujimori}}, \bibinfo {author}
  {\bibfnamefont {A.}~\bibnamefont {Fujimori}}, \bibinfo {author}
  {\bibfnamefont {T.}~\bibnamefont {Mizokawa}}, \bibinfo {author}
  {\bibfnamefont {Y.}~\bibnamefont {Yasui}}, \bibinfo {author} {\bibfnamefont
  {T.}~\bibnamefont {Nishikawa}}, \ and\ \bibinfo {author} {\bibfnamefont
  {M.}~\bibnamefont {Sato}},\ }\href {\doibase 10.1103/PhysRevB.69.195116}
  {\bibfield  {journal} {\bibinfo  {journal} {Phys. Rev. B}\ }\textbf {\bibinfo
  {volume} {69}},\ \bibinfo {pages} {195116} (\bibinfo {year}
  {2004})}\BibitemShut {NoStop}%
\bibitem [{\citenamefont {Yasui}\ \emph {et~al.}(1995)\citenamefont {Yasui},
  \citenamefont {Nishikawa}, \citenamefont {Kobayashi}, \citenamefont {Sato},
  \citenamefont {Nishioka},\ and\ \citenamefont {Kontani}}]{Yasui1995}%
  \BibitemOpen
  \bibfield  {author} {\bibinfo {author} {\bibfnamefont {Y.}~\bibnamefont
  {Yasui}}, \bibinfo {author} {\bibfnamefont {T.}~\bibnamefont {Nishikawa}},
  \bibinfo {author} {\bibfnamefont {Y.}~\bibnamefont {Kobayashi}}, \bibinfo
  {author} {\bibfnamefont {M.}~\bibnamefont {Sato}}, \bibinfo {author}
  {\bibfnamefont {T.}~\bibnamefont {Nishioka}}, \ and\ \bibinfo {author}
  {\bibfnamefont {M.}~\bibnamefont {Kontani}},\ }\href {\doibase
  10.1143/JPSJ.64.3890} {\bibfield  {journal} {\bibinfo  {journal} {J. Phys.
  Soc. Jpn.}\ }\textbf {\bibinfo {volume} {64}},\ \bibinfo {pages} {3890}
  (\bibinfo {year} {1995})}\BibitemShut {NoStop}%
\bibitem [{\citenamefont {Bruker Nano~GmbH}(20xx)}]{esprit}%
  \BibitemOpen
  \bibfield  {author} {\bibinfo {author} {\bibfnamefont {G.}~\bibnamefont
  {Bruker Nano~GmbH}, \bibfnamefont {Berlin}},\ }\href@noop {} {\enquote
  {\bibinfo {title} {Software: Esprit family},}\ } (\bibinfo {year}
  {20xx})\BibitemShut {NoStop}%
\bibitem [{\citenamefont {Pet\v{r}\'{i}\v{c}ek}\ \emph
  {et~al.}(2014)\citenamefont {Pet\v{r}\'{i}\v{c}ek}, \citenamefont
  {Du\v{s}ek},\ and\ \citenamefont {Palatinus}}]{jana2006}%
  \BibitemOpen
  \bibfield  {author} {\bibinfo {author} {\bibfnamefont {V.}~\bibnamefont
  {Pet\v{r}\'{i}\v{c}ek}}, \bibinfo {author} {\bibfnamefont {M.}~\bibnamefont
  {Du\v{s}ek}}, \ and\ \bibinfo {author} {\bibfnamefont {L.}~\bibnamefont
  {Palatinus}},\ }\href {\doibase 10.1515/zkri-2014-1737} {\bibfield  {journal}
  {\bibinfo  {journal} {Z. Kristallogr. Cryst. Mater.}\ }\textbf {\bibinfo
  {volume} {229}},\ \bibinfo {pages} {345–352} (\bibinfo {year}
  {2014})}\BibitemShut {NoStop}%
\bibitem [{\citenamefont {Palatinus}\ and\ \citenamefont
  {Chapuis}(2007)}]{superflip}%
  \BibitemOpen
  \bibfield  {author} {\bibinfo {author} {\bibfnamefont {L.}~\bibnamefont
  {Palatinus}}\ and\ \bibinfo {author} {\bibfnamefont {G.}~\bibnamefont
  {Chapuis}},\ }\href {\doibase 10.1107/s0021889807029238} {\bibfield
  {journal} {\bibinfo  {journal} {J. Appl. Crystallogr.}\ }\textbf {\bibinfo
  {volume} {40}},\ \bibinfo {pages} {786} (\bibinfo {year} {2007})}\BibitemShut
  {NoStop}%
\bibitem [{\citenamefont {Momma}\ and\ \citenamefont {Izumi}(2011)}]{vesta}%
  \BibitemOpen
  \bibfield  {author} {\bibinfo {author} {\bibfnamefont {K.}~\bibnamefont
  {Momma}}\ and\ \bibinfo {author} {\bibfnamefont {F.}~\bibnamefont {Izumi}},\
  }\href {\doibase 10.1107/S0021889811038970} {\bibfield  {journal} {\bibinfo
  {journal} {J. of Appl. Crystallogr.}\ }\textbf {\bibinfo {volume} {44}},\
  \bibinfo {pages} {1272} (\bibinfo {year} {2011})}\BibitemShut {NoStop}%
\bibitem [{\citenamefont {Ettema}\ and\ \citenamefont
  {Haas}(1993)}]{Ettema1993}%
  \BibitemOpen
  \bibfield  {author} {\bibinfo {author} {\bibfnamefont {A.~R. H.~F.}\
  \bibnamefont {Ettema}}\ and\ \bibinfo {author} {\bibfnamefont
  {C.}~\bibnamefont {Haas}},\ }\href {\doibase 10.1088/0953-8984/5/23/008}
  {\bibfield  {journal} {\bibinfo  {journal} {J. Condens. Matter Phys.}\
  }\textbf {\bibinfo {volume} {5}},\ \bibinfo {pages} {3817} (\bibinfo {year}
  {1993})}\BibitemShut {NoStop}%
\bibitem [{\citenamefont {Brandt}(2003)}]{Brandt2003}%
  \BibitemOpen
  \bibfield  {author} {\bibinfo {author} {\bibfnamefont {J.}~\bibnamefont
  {Brandt}},\ }\emph {\bibinfo {title} {Geometric and electronic structure of
  misﬁt layered compounds and epitaxial thin ﬁlms of PbS on transition
  metal dichalcogenides}},\ \href@noop {} {Ph.D. thesis},\ \bibinfo  {school}
  {Mathematisch-Naturwissenschaftlichen Fakult\'{a}t,
  Christian-Albrechts-Universit{\"a}t, Kiel} (\bibinfo {year}
  {2003})\BibitemShut {NoStop}%
\bibitem [{\citenamefont {Hossain}\ \emph {et~al.}(2018)\citenamefont
  {Hossain}, \citenamefont {Wu}, \citenamefont {Wen}, \citenamefont {Liu},
  \citenamefont {Wang},\ and\ \citenamefont {Xie}}]{Hossain2018}%
  \BibitemOpen
  \bibfield  {author} {\bibinfo {author} {\bibfnamefont {M.}~\bibnamefont
  {Hossain}}, \bibinfo {author} {\bibfnamefont {J.}~\bibnamefont {Wu}},
  \bibinfo {author} {\bibfnamefont {W.}~\bibnamefont {Wen}}, \bibinfo {author}
  {\bibfnamefont {H.}~\bibnamefont {Liu}}, \bibinfo {author} {\bibfnamefont
  {X.}~\bibnamefont {Wang}}, \ and\ \bibinfo {author} {\bibfnamefont
  {L.}~\bibnamefont {Xie}},\ }\href {\doibase 10.1002/admi.201800528}
  {\bibfield  {journal} {\bibinfo  {journal} {Adv. Mater. Interfaces}\ }\textbf
  {\bibinfo {volume} {5}},\ \bibinfo {pages} {1800528} (\bibinfo {year}
  {2018})}\BibitemShut {NoStop}%
\bibitem [{\citenamefont {Zhang}\ \emph {et~al.}(2017)\citenamefont {Zhang},
  \citenamefont {Niu}, \citenamefont {Yang}, \citenamefont {Gong},
  \citenamefont {Ji}, \citenamefont {Shi}, \citenamefont {Fang}, \citenamefont
  {Jiang}, \citenamefont {Li}, \citenamefont {Zhou}, \citenamefont {Gu},
  \citenamefont {Wu},\ and\ \citenamefont {Zhang}}]{Zhang2017}%
  \BibitemOpen
  \bibfield  {author} {\bibinfo {author} {\bibfnamefont {Z.}~\bibnamefont
  {Zhang}}, \bibinfo {author} {\bibfnamefont {J.}~\bibnamefont {Niu}}, \bibinfo
  {author} {\bibfnamefont {P.}~\bibnamefont {Yang}}, \bibinfo {author}
  {\bibfnamefont {Y.}~\bibnamefont {Gong}}, \bibinfo {author} {\bibfnamefont
  {Q.}~\bibnamefont {Ji}}, \bibinfo {author} {\bibfnamefont {J.}~\bibnamefont
  {Shi}}, \bibinfo {author} {\bibfnamefont {Q.}~\bibnamefont {Fang}}, \bibinfo
  {author} {\bibfnamefont {S.}~\bibnamefont {Jiang}}, \bibinfo {author}
  {\bibfnamefont {H.}~\bibnamefont {Li}}, \bibinfo {author} {\bibfnamefont
  {X.}~\bibnamefont {Zhou}}, \bibinfo {author} {\bibfnamefont {L.}~\bibnamefont
  {Gu}}, \bibinfo {author} {\bibfnamefont {X.}~\bibnamefont {Wu}}, \ and\
  \bibinfo {author} {\bibfnamefont {Y.}~\bibnamefont {Zhang}},\ }\href
  {\doibase https://doi.org/10.1002/adma.201702359} {\bibfield  {journal}
  {\bibinfo  {journal} {Adv. Mater.}\ }\textbf {\bibinfo {volume} {29}},\
  \bibinfo {pages} {1702359} (\bibinfo {year} {2017})}\BibitemShut {NoStop}%
\bibitem [{\citenamefont {Hossain}\ \emph {et~al.}(2021)\citenamefont
  {Hossain}, \citenamefont {Iqbal}, \citenamefont {Wu},\ and\ \citenamefont
  {Xie}}]{Hossain2021}%
  \BibitemOpen
  \bibfield  {author} {\bibinfo {author} {\bibfnamefont {M.}~\bibnamefont
  {Hossain}}, \bibinfo {author} {\bibfnamefont {M.}~\bibnamefont {Iqbal}},
  \bibinfo {author} {\bibfnamefont {J.}~\bibnamefont {Wu}}, \ and\ \bibinfo
  {author} {\bibfnamefont {L.}~\bibnamefont {Xie}},\ }\href {\doibase
  10.1039/D0RA07868A} {\bibfield  {journal} {\bibinfo  {journal} {RSC Adv.}\
  }\textbf {\bibinfo {volume} {11}},\ \bibinfo {pages} {2624} (\bibinfo {year}
  {2021})}\BibitemShut {NoStop}%
\bibitem [{\citenamefont {Silversmit}\ \emph {et~al.}(2004)\citenamefont
  {Silversmit}, \citenamefont {Depla}, \citenamefont {Poelman}, \citenamefont
  {Marin},\ and\ \citenamefont {{De Gryse}}}]{Silversmit2004}%
  \BibitemOpen
  \bibfield  {author} {\bibinfo {author} {\bibfnamefont {G.}~\bibnamefont
  {Silversmit}}, \bibinfo {author} {\bibfnamefont {D.}~\bibnamefont {Depla}},
  \bibinfo {author} {\bibfnamefont {H.}~\bibnamefont {Poelman}}, \bibinfo
  {author} {\bibfnamefont {G.~B.}\ \bibnamefont {Marin}}, \ and\ \bibinfo
  {author} {\bibfnamefont {R.}~\bibnamefont {{De Gryse}}},\ }\href {\doibase
  10.1016/j.elspec.2004.03.004} {\bibfield  {journal} {\bibinfo  {journal} {J.
  Electron Spectrosc. Relat. Phenom.}\ }\textbf {\bibinfo {volume} {135}},\
  \bibinfo {pages} {167} (\bibinfo {year} {2004})}\BibitemShut {NoStop}%
\bibitem [{\citenamefont {Göhler}\ \emph {et~al.}(2022)\citenamefont
  {Göhler}, \citenamefont {Ramasubramanian}, \citenamefont {Rajak},
  \citenamefont {Rösch}, \citenamefont {Schütze}, \citenamefont {Wolff},
  \citenamefont {Cordova}, \citenamefont {Johnson},\ and\ \citenamefont
  {Seyller}}]{Goehler2022}%
  \BibitemOpen
  \bibfield  {author} {\bibinfo {author} {\bibfnamefont {F.}~\bibnamefont
  {Göhler}}, \bibinfo {author} {\bibfnamefont {S.}~\bibnamefont
  {Ramasubramanian}}, \bibinfo {author} {\bibfnamefont {S.~K.}\ \bibnamefont
  {Rajak}}, \bibinfo {author} {\bibfnamefont {N.}~\bibnamefont {Rösch}},
  \bibinfo {author} {\bibfnamefont {A.}~\bibnamefont {Schütze}}, \bibinfo
  {author} {\bibfnamefont {S.}~\bibnamefont {Wolff}}, \bibinfo {author}
  {\bibfnamefont {D.~L.~M.}\ \bibnamefont {Cordova}}, \bibinfo {author}
  {\bibfnamefont {D.~C.}\ \bibnamefont {Johnson}}, \ and\ \bibinfo {author}
  {\bibfnamefont {T.}~\bibnamefont {Seyller}},\ }\href {\doibase
  10.1039/D2NR01071B} {\bibfield  {journal} {\bibinfo  {journal} {Nanoscale}\
  }\textbf {\bibinfo {volume} {14}},\ \bibinfo {pages} {10143} (\bibinfo {year}
  {2022})}\BibitemShut {NoStop}%
\bibitem [{\citenamefont {Göhler}\ \emph {et~al.}(2018)\citenamefont
  {Göhler}, \citenamefont {Mitchson}, \citenamefont {Alemayehu}, \citenamefont
  {Speck}, \citenamefont {Wanke}, \citenamefont {Johnson},\ and\ \citenamefont
  {Seyller}}]{Goehler2018}%
  \BibitemOpen
  \bibfield  {author} {\bibinfo {author} {\bibfnamefont {F.}~\bibnamefont
  {Göhler}}, \bibinfo {author} {\bibfnamefont {G.}~\bibnamefont {Mitchson}},
  \bibinfo {author} {\bibfnamefont {M.~B.}\ \bibnamefont {Alemayehu}}, \bibinfo
  {author} {\bibfnamefont {F.}~\bibnamefont {Speck}}, \bibinfo {author}
  {\bibfnamefont {M.}~\bibnamefont {Wanke}}, \bibinfo {author} {\bibfnamefont
  {D.~C.}\ \bibnamefont {Johnson}}, \ and\ \bibinfo {author} {\bibfnamefont
  {T.}~\bibnamefont {Seyller}},\ }\href {\doibase 10.1088/1361-648X/aaa212}
  {\bibfield  {journal} {\bibinfo  {journal} {Journal of Physics: Condensed
  Matter}\ }\textbf {\bibinfo {volume} {30}},\ \bibinfo {pages} {055001}
  (\bibinfo {year} {2018})}\BibitemShut {NoStop}%
\bibitem [{\citenamefont {Leiro}\ \emph {et~al.}(1998)\citenamefont {Leiro},
  \citenamefont {Laajalehto}, \citenamefont {Kartio},\ and\ \citenamefont
  {Heinonen}}]{LEIRO1998}%
  \BibitemOpen
  \bibfield  {author} {\bibinfo {author} {\bibfnamefont {J.}~\bibnamefont
  {Leiro}}, \bibinfo {author} {\bibfnamefont {K.}~\bibnamefont {Laajalehto}},
  \bibinfo {author} {\bibfnamefont {I.}~\bibnamefont {Kartio}}, \ and\ \bibinfo
  {author} {\bibfnamefont {M.}~\bibnamefont {Heinonen}},\ }\href {\doibase
  https://doi.org/10.1016/S0039-6028(98)00479-8} {\bibfield  {journal}
  {\bibinfo  {journal} {Surface Science}\ }\textbf {\bibinfo {volume}
  {412-413}},\ \bibinfo {pages} {L918} (\bibinfo {year} {1998})}\BibitemShut
  {NoStop}%
\bibitem [{\citenamefont {Fong}\ \emph {et~al.}(2006)\citenamefont {Fong},
  \citenamefont {Kolpak}, \citenamefont {Eastman}, \citenamefont {Streiffer},
  \citenamefont {Fuoss}, \citenamefont {Stephenson}, \citenamefont {Thompson},
  \citenamefont {Kim}, \citenamefont {Choi}, \citenamefont {Eom}, \citenamefont
  {Grinberg},\ and\ \citenamefont {Rappe}}]{Fong2006}%
  \BibitemOpen
  \bibfield  {author} {\bibinfo {author} {\bibfnamefont {D.~D.}\ \bibnamefont
  {Fong}}, \bibinfo {author} {\bibfnamefont {A.~M.}\ \bibnamefont {Kolpak}},
  \bibinfo {author} {\bibfnamefont {J.~A.}\ \bibnamefont {Eastman}}, \bibinfo
  {author} {\bibfnamefont {S.~K.}\ \bibnamefont {Streiffer}}, \bibinfo {author}
  {\bibfnamefont {P.~H.}\ \bibnamefont {Fuoss}}, \bibinfo {author}
  {\bibfnamefont {G.~B.}\ \bibnamefont {Stephenson}}, \bibinfo {author}
  {\bibfnamefont {C.}~\bibnamefont {Thompson}}, \bibinfo {author}
  {\bibfnamefont {D.~M.}\ \bibnamefont {Kim}}, \bibinfo {author} {\bibfnamefont
  {K.~J.}\ \bibnamefont {Choi}}, \bibinfo {author} {\bibfnamefont {C.~B.}\
  \bibnamefont {Eom}}, \bibinfo {author} {\bibfnamefont {I.}~\bibnamefont
  {Grinberg}}, \ and\ \bibinfo {author} {\bibfnamefont {A.~M.}\ \bibnamefont
  {Rappe}},\ }\href {\doibase 10.1103/PhysRevLett.96.127601} {\bibfield
  {journal} {\bibinfo  {journal} {Phys. Rev. Lett.}\ }\textbf {\bibinfo
  {volume} {96}},\ \bibinfo {pages} {127601} (\bibinfo {year}
  {2006})}\BibitemShut {NoStop}%
\bibitem [{\citenamefont {Baniecki}\ \emph {et~al.}(2009)\citenamefont
  {Baniecki}, \citenamefont {Ishii}, \citenamefont {Kurihara}, \citenamefont
  {Yamanaka}, \citenamefont {Yano}, \citenamefont {Shinozaki}, \citenamefont
  {Imada},\ and\ \citenamefont {Kobayashi}}]{Baniecki2009}%
  \BibitemOpen
  \bibfield  {author} {\bibinfo {author} {\bibfnamefont {J.}~\bibnamefont
  {Baniecki}}, \bibinfo {author} {\bibfnamefont {M.}~\bibnamefont {Ishii}},
  \bibinfo {author} {\bibfnamefont {K.}~\bibnamefont {Kurihara}}, \bibinfo
  {author} {\bibfnamefont {K.}~\bibnamefont {Yamanaka}}, \bibinfo {author}
  {\bibfnamefont {T.}~\bibnamefont {Yano}}, \bibinfo {author} {\bibfnamefont
  {K.}~\bibnamefont {Shinozaki}}, \bibinfo {author} {\bibfnamefont
  {T.}~\bibnamefont {Imada}}, \ and\ \bibinfo {author} {\bibfnamefont
  {Y.}~\bibnamefont {Kobayashi}},\ }\href {\doibase 10.1063/1.3169654}
  {\bibfield  {journal} {\bibinfo  {journal} {J. Appl. Phys.}\ }\textbf
  {\bibinfo {volume} {106}},\ \bibinfo {pages} {054109} (\bibinfo {year}
  {2009})}\BibitemShut {NoStop}%
\bibitem [{\citenamefont {Krug}\ \emph {et~al.}(2010)\citenamefont {Krug},
  \citenamefont {Barrett}, \citenamefont {Petraru}, \citenamefont {Locatelli},
  \citenamefont {Mentes}, \citenamefont {Ni{\~n}o.}, \citenamefont
  {Rahmanizadeh}, \citenamefont {Bihlmayer},\ and\ \citenamefont
  {Schneider}}]{Krug2010}%
  \BibitemOpen
  \bibfield  {author} {\bibinfo {author} {\bibfnamefont {I.}~\bibnamefont
  {Krug}}, \bibinfo {author} {\bibfnamefont {N.}~\bibnamefont {Barrett}},
  \bibinfo {author} {\bibfnamefont {A.}~\bibnamefont {Petraru}}, \bibinfo
  {author} {\bibfnamefont {A.}~\bibnamefont {Locatelli}}, \bibinfo {author}
  {\bibfnamefont {T.}~\bibnamefont {Mentes}}, \bibinfo {author} {\bibfnamefont
  {M.}~\bibnamefont {Ni{\~n}o.}}, \bibinfo {author} {\bibfnamefont
  {K.}~\bibnamefont {Rahmanizadeh}}, \bibinfo {author} {\bibfnamefont
  {G.}~\bibnamefont {Bihlmayer}}, \ and\ \bibinfo {author} {\bibfnamefont
  {C.}~\bibnamefont {Schneider}},\ }\href {\doibase 10.1063/1.3523359}
  {\bibfield  {journal} {\bibinfo  {journal} {Appl. Phys. Lett.}\ }\textbf
  {\bibinfo {volume} {97}},\ \bibinfo {pages} {222903} (\bibinfo {year}
  {2010})}\BibitemShut {NoStop}%
\end{thebibliography}
\end{document}

\title[Sliding ferroelectricity in a bulk misfit layer compound (PbS)$_{1.18}$VS$_2$]{Supplemental Material\\Sliding ferroelectricity in a bulk misfit layer compound (PbS)$_{1.18}$VS$_2$}% Force line breaks with \\

\author{Cinthia Antunes Corr\^{e}a}
\affiliation{FZU - Institute of Physics of the Czech Academy of Sciences, Na Slovance 2, Praha 8, CZ-18 221, Czech Republic}
\affiliation{Department of Physics of Materials, Faculty of Mathematics and Physics, Charles University, Ke Karlovu 3, CZ-121 16, Prague 2, Czech Republic}

\author{Ji\v{r}\'{i} Voln\'{y}}
\author{Kate\v{r}ina Tetalov\'{a}}
\author{Kl\'{a}ra Uhl\'{i}\v{r}ov\'{a}}
\affiliation{Department of Condensed Matter Physics, Faculty of Mathematics and Physics, Charles University, Ke Karlovu 3, CZ-121 16, Prague 2, Czech Republic}

\author{V\'{a}clav Pet\v{r}\'{i}\v{c}ek}
\affiliation{FZU - Institute of Physics of the Czech Academy of Sciences, Na Slovance 2, Praha 8, CZ-18 221, Czech Republic}

\author{Martin Vondr\'{a}\v{c}ek}
\author{Jan Honolka}
\affiliation{FZU - Institute of Physics of the Czech Academy of Sciences, Na Slovance 2, Praha 8, CZ-18 221, Czech Republic}

\author{Tim Verhagen }
\email{verhagen@fzu.cz}
\affiliation{FZU - Institute of Physics of the Czech Academy of Sciences, Na Slovance 2, Praha 8, CZ-18 221, Czech Republic}
\affiliation{Institute of Physics, Faculty of Mathematics and Physics, Charles University, Ke Karlovu 3, CZ-121 16, Prague 2, Czech Republic}

\maketitle

\section{Methods}
Single crystals of (PbS)$_{1.18}$VS$_2$ were grown using chemical vapor transport according to the modified recipe by Gotoh et al.~\cite{Gotoh1990}.  The pure elements (99,5\% vanadium powder, small pieces of 99.999\% lead and 99.999\% sulphur, all Alfa Aesar) with molar composition  VPb$_{1.12}$S$_{3.12}$ were weighted and loaded in an approximately 25~cm long fused silica tube in Ar filled glove-box and subsequently sealed in vacuum of $10^{-6}$~mbar. 

The sample was first slowly heated in a vertical position up to 720$^{\circ}C$ and then placed into a two-zone horizontal furnace. The growth temperatures were 720$^{\circ}C$ and 650$^{\circ}C$ in the hot and cold zone, respectively. The growth process took 1~-~6 weeks, depending on the desired size of the grown single crystals. Smaller crystals were grown for the XRD measurements. For the growth, we used about 3 grams of materials. However, only a small part of it was transferred into single crystals. When handled in a glove-box, the remaining polycrystalline material can be reused for another growth.    

\begin{figure}
 \includegraphics[width=\columnwidth] {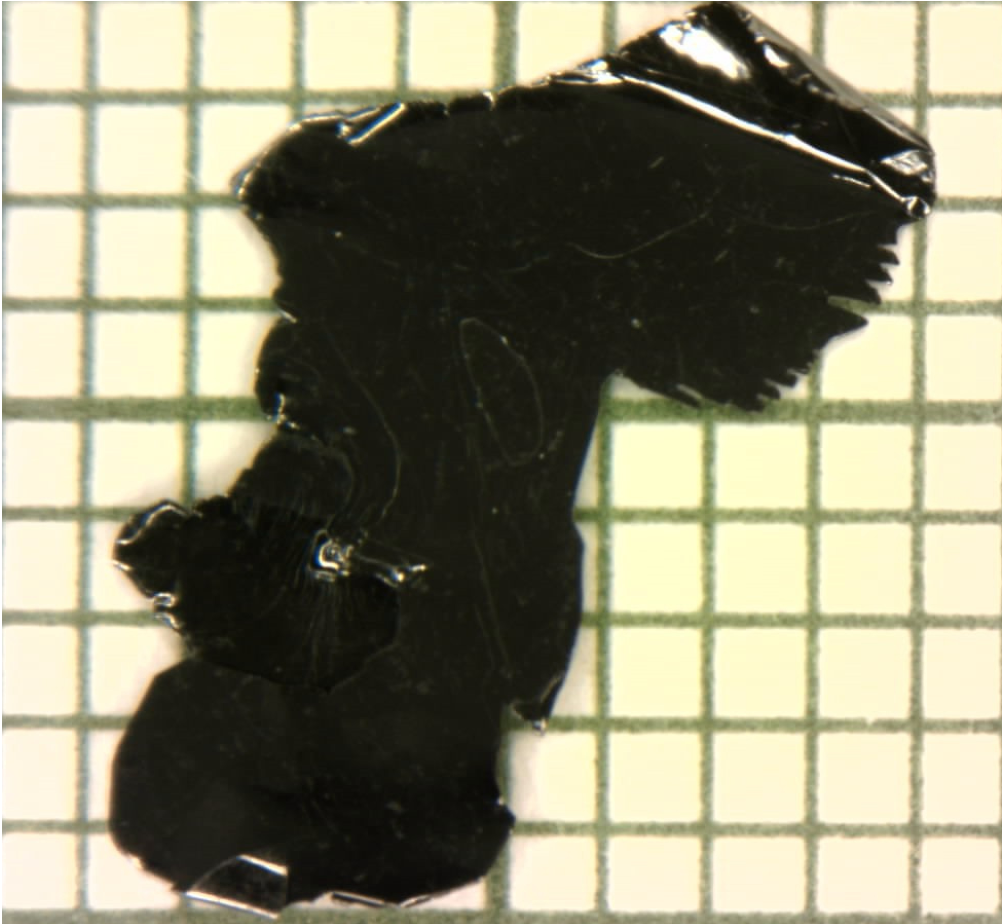}
  \caption{Optical microscope image of typical (PbS)VS$_2$ crystal on millimeter paper. The sample thickness is of the order of a few tens of $\mu$m and the c-axis is pointing perpendicular to the plane.}
  \label{fgr:fig_s1}
\end{figure}

The as-grown single crystals (typical sample is shown in Fig.~\ref{fgr:fig_s1}) are shiny, flat, relatively large (a couple of mm$^2$), and very thin (few tens of $\mu$m), with the \textit{c}-axis perpendicular to the surface.

It was demonstrated that secondary electrons (SE) in a scanning electron microscope (SEM) can be used for imaging ferroelectric domains. \cite{LEBIHAN1972, Hunnestad2020} and recently, it was shown ~\cite{Andersen2021}, secondary electron channeling contrast can be used for imaging of the moir\'{e} domains in manually twisted few-layer van der Waals materials. In (PbS)$_{1.18}$VS$_2$ crystals, the SE contrast has been obtained without special effort at various electron energies from 5 to 20~keV with the electron beam perpendicular to the sample surface.  

The composition of the selected crystals was determined using energy-dispersive x-ray spectroscopy (EDS) and X-ray photoelectron spectroscopy (XPS). EDS was performed on a Tescan Mira LMH SEM equipped with an energy-dispersive x-ray detector (EDX) Bruker AXS utilizing ESPRIT software~\cite{esprit} (based on a non-standard method). The resulting stoichiometry was reproducible amongst samples across different batches, confirming the presence of (PbS)$_{1.18}$VS$_2$, however, due to the overlap of the lead M- and sulfur K-shell peaks, the accuracy is rather low (with the relative error of 10~\%).

XPS and elemental mapping measurements were carried out using an Omicron NanoESCA instrument with monochromatized Al K$\alpha$ radiation (1486.7~eV). Ultraviolet photoelectron spectroscopy (UPS) and Photoemission electron microscopy (PEEM) were performed using a helium discharge lamp at h$\nu$ = 21.2~eV.

Single-crystal x-ray diffraction was performed at 95 K using CuK$\alpha$ radiation ($\lambda=1.54$~\AA ) on a SuperNova diffractometer, equipped with a sealed microfocus x-ray tube, a mirror collimator, and an Atlas S2 CCD. Diffraction data of both subsystems of the composite and the satellites were integrated by importing the DCRED file generated by Jana2006~\cite{jana2006} to CrysAlisPro~\cite{crysalis}, using numeric absorption correction with a multifaceted crystal combined with spherical harmonics for the empirical absorption correction. We used the program CrysAlisPro and Vesta~\cite{vesta} for the images.

Using a gallium-focussed ion beam (FIB) Zeiss Auriga Compact scanning electron microscope, a lamella was milled out perpendicular to the crystal surface i.e. parallel with the \textit{c}-axis. STEM images were acquired using an FEI Titan Themis cubed transmission electron microscope (TEM) operated at 300 kV.

The topography and domain structure was investigated using a Bruker Multimode 8 atomic force microscopy (AFM) system using tapping mode and electrostatic force microscopy (EFM) with conducting probes ($k = 2.8$ N/m, resonance frequency of 80 kHz). The lift height for the EFM was typically 15-20~nm, when increasing the lift height, the phase contrast quickly decreased.

\begin{figure}
 \includegraphics[width=\columnwidth] {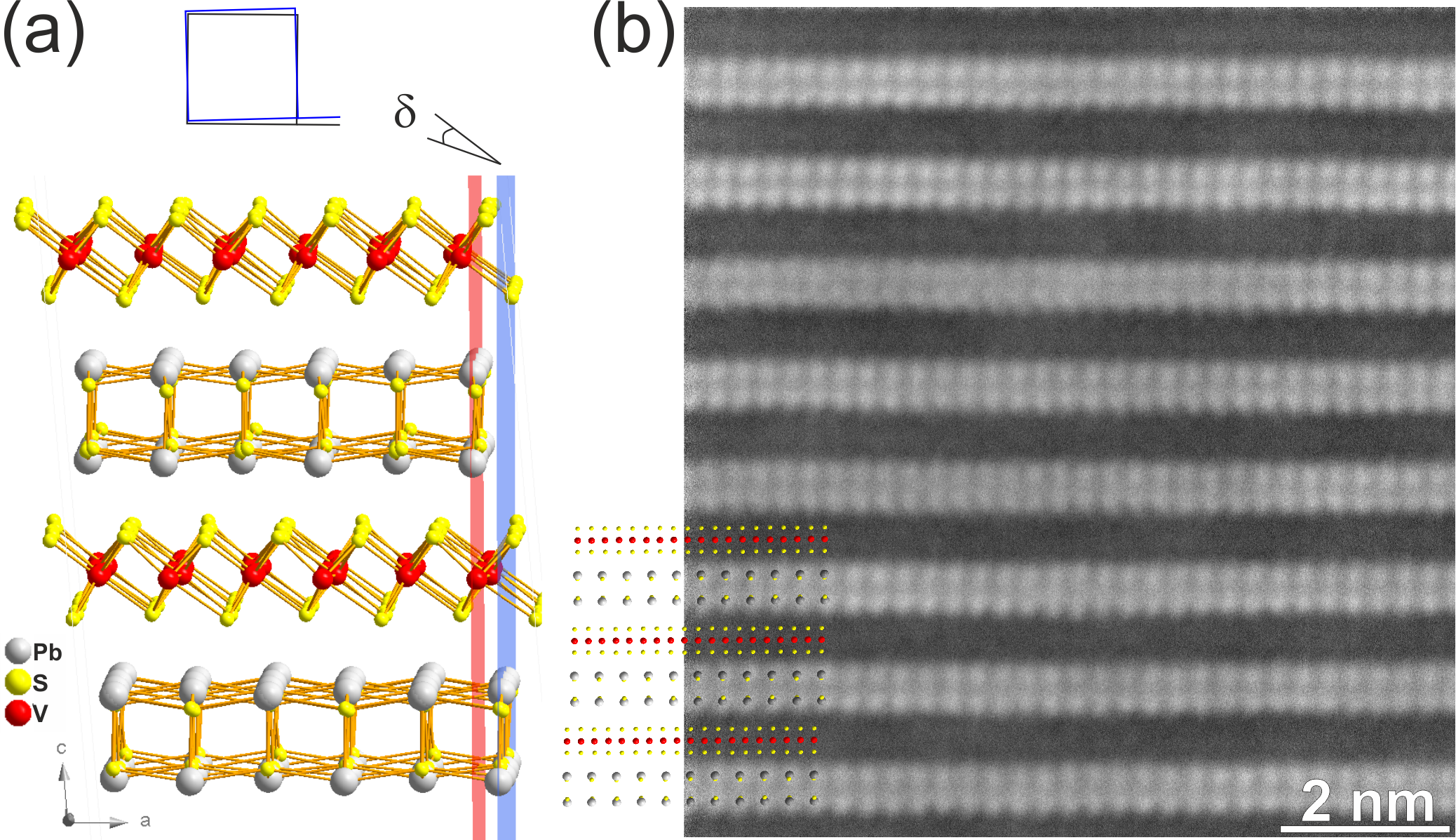}
  \caption{Layered structure of the crystal. (a) Schematic overview of the rotation between two PbS layers. Blue and black squares represent the top view of two PbS layers, where the rotation between the two planes is smaller than $\delta\simeq 1.2\degree$. The blue plane goes through the right-most Pb atoms of the bottom PbS layer, while the red plane goes through the Pb atoms of the subsequent PbS layer. (b) HAADF-STEM image with overlapped atomic types.}
  \label{fgr:fig_s2}
\end{figure}

\section{XRD}
All the single crystals that were measured, contained several twins due to the layer composite structure, which allows for various layer configurations. Figure~\ref{fgr:fig_s2}(a) shows a schematic of the rotation between two PbS layers, while the HAADF-STEM cross-section of the crystal (Fig.~\ref{fgr:fig_s2}(b)) shows the layered structure with superposed atomic layers.

\begin{figure}
 \includegraphics[width=\columnwidth] {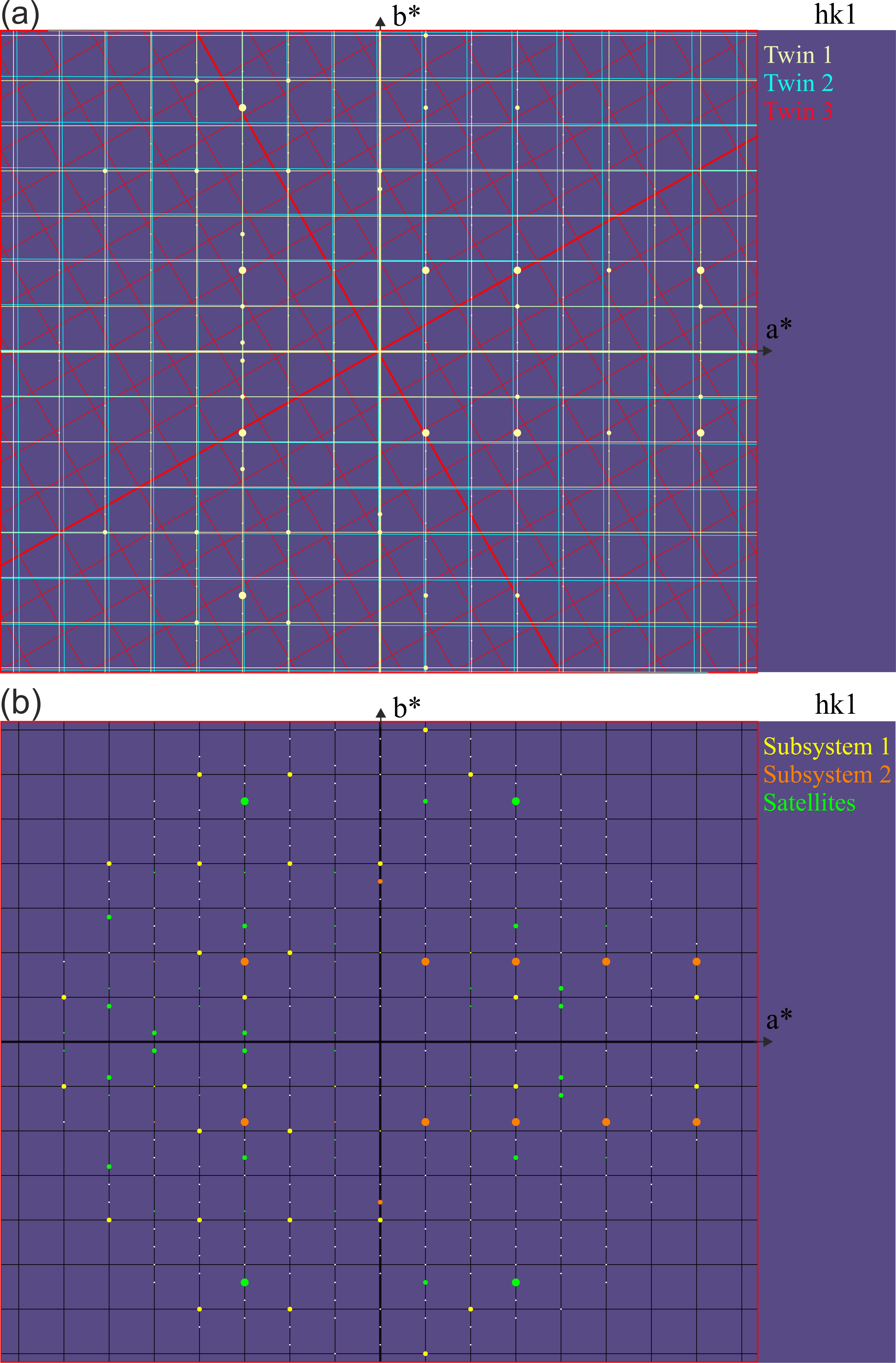}
  \caption{Reciprocal space of (PbS)$_{1.18}$VS$_2$ in Jana2020. (a) Section $hk1$: three of the possible twins observed in the data. Twin 1 (in light yellow) includes the main reflections of subsystem 1, the main reflections of subsystem 2, and satellite reflections. Twins 2 and 3 are shown in light blue and red, respectively. (b) Twin 1: the reflections of subsystem 1 are in yellow, those from subsystem 2 are in orange, and satellite reflections are in green.}
  \label{fgr:fig_s3}
\end{figure}

Figure~\ref{fgr:fig_s3}(a) shows section \textit{hk1} of the reciprocal space of (PbS)$_{1.18}$VS$_2$ containing the twin 1 (subsystem 1, subsystem 2, and satellites) in light yellow, twin 2 in light blue, and twin 3 in red. Twin 1 is the component with the highest percentage of indexed reflections.
Figure~\ref{fgr:fig_s3}(b) contains only twin 1 and shows the main reflections from subsystem 1 in yellow, the main reflections belonging to subsystem 2 in orange, and satellites are shown in green.

In Tab.~\ref{tab:tab_s1}, the twins observed in different (PbS)$_{1.18}$VS$_2$ crystals are shown, where the twist angle from the dominant twin is set as a reference to 0$\degree$. As clearly visible, the most occurring twist angle from the twins is close to $\pm$180$\degree$. It should be noted, that from the single crystal XRD data, the stacking order between the twins cannot be determined.

\begin{longtable*}{ccc}
\caption{Twins present different (PbS)$_{1.18}$VS$_2$ crystals, their twist angle along the $c$-axis with respect to the twin with the largest presence and, if present, their rotation along the a and b and c axes. } \label{tab:tab_s1} \\

\toprule
\multicolumn{1}{c}{Twin} & \multicolumn{1}{c}{Twist angle c axis ($\degree$)} & \multicolumn{1}{c}{Rotation around} \\ 
\toprule
\endfirsthead

\multicolumn{3}{c}%
{{\bfseries \tablename\ \thetable{} -- continued from previous page}} \\
\toprule
\multicolumn{1}{c}{Twin} & \multicolumn{1}{c}{Twist angle c-axis ($\degree$)} & \multicolumn{1}{c}{Rotation along planes} \\ 
\toprule
\endhead

\hline \multicolumn{3}{c}{{Continued on next page}} \\ \hline
\endfoot

\bottomrule
\endlastfoot

\textbf{Crystal 1} \\
1 &  0     &  \\
2 &  180.0 & a,b \\
3 & -180.0 & a,b \\
4 & -179.9 & a,b \\
5 & 0.1    & a,b,c \\
\textbf{Crystal 2} \\
1 &  0     &  \\
2 & -179.7 & b \\
3 & 37.3   & b,c \\
4 & 124.0  & a,b \\
5 & 118.5  & a \\
\textbf{Crystal 3} \\
1 &  0     &  \\
2 & 7.9    & a,b \\
3 & -179.8 & a,b \\
4 & -175.5 & b \\
5 & 179.2  & a,b,c \\
\textbf{Crystal 4} \\
1 &  0     &  \\
2 & -91.2  & a,c \\
3 & -179.9 & a,b \\
4 & 176.8  & a,b \\
5 & 6.4    & a,b \\
\textbf{Crystal 5} \\
1 &  0     &  \\
2 & -180.0 & a,b \\
3 & 78.8   & a,c \\
4 & -179.9 & a,b \\
5 & 31.7   & a,c \\
\textbf{Crystal 6} \\
1 &  0     &  \\
2 & -119.2 & a,c \\
3 & -180.0 & a,b \\
4 & -179.9 & a,c \\
5 & -179.9 & a,b \\
\textbf{Crystal 7} \\
1 &  0     &  \\
2 & -180.0 & a,b \\
3 & -180.0 & a,b \\
4 & 180.0  & a,b \\
5 & 93.2  &  a,b,c\\
\textbf{Crystal 8} \\
1 &  0     &  \\
2 & 179.9  &  a,b\\
3 & 4.3    &  a,b,c\\
4 & -116.1 &  a,b,c\\
5 & 5.1    &  a,b,c\\
\textbf{Crystal 9} \\
1 &  0      &  \\
2 & 180.0   &  a,b\\
3 & -162.1  &  a,c\\
4 & 89.9    &  a,c\\
5 & -168.5  &  a,b\\
\textbf{Crystal 10} \\
1 &  0     &  \\
2 & -180   &  a,c\\
3 & 173.2  &  a,b\\
4 & 119.69 &  a,c\\
5 & -179.9 &  a,c\\
\textbf{Crystal 11} \\
1 &  0     &  \\
2 & -121.1 &  a,c\\
3 & 119.5  &  a,c\\
\end{longtable*}

\section{SEM}
\begin{figure}
 \includegraphics[width=\columnwidth] {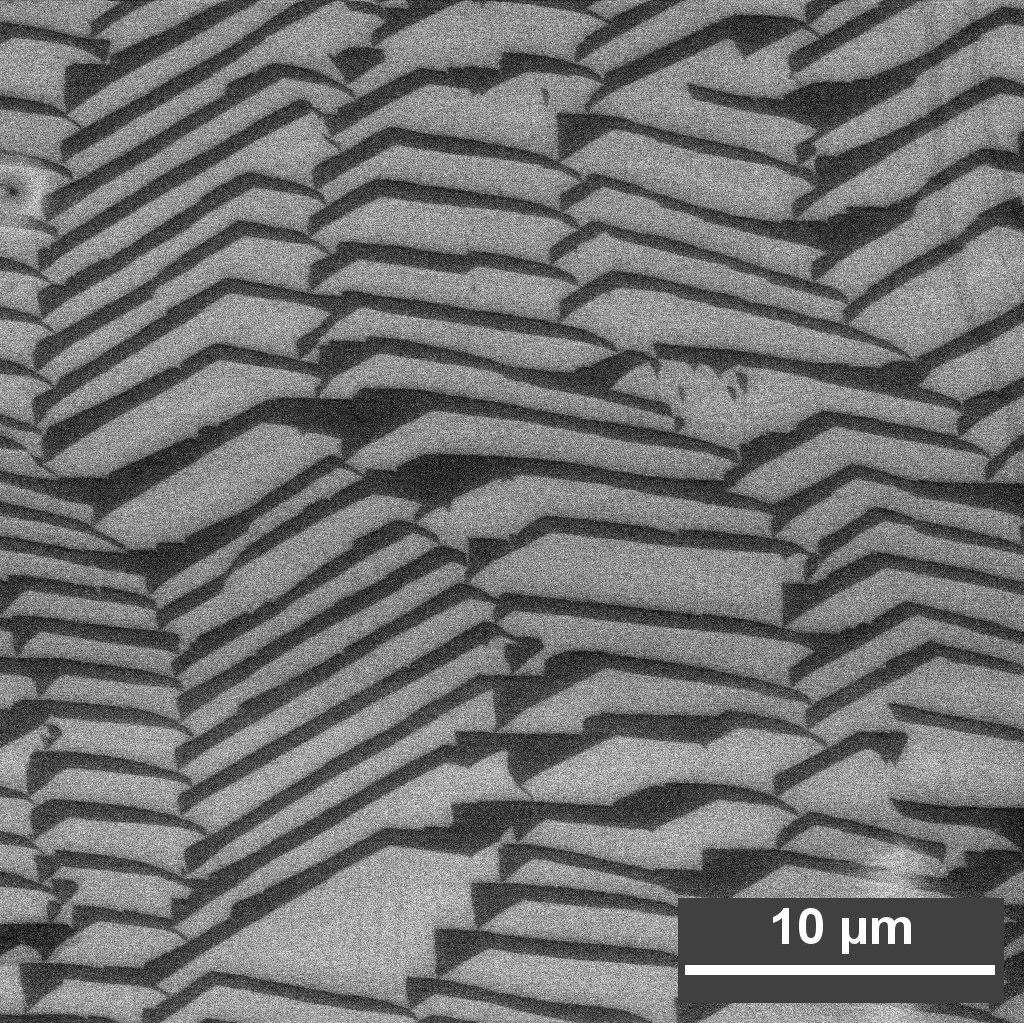}
  \caption{SE image of a cleaved (PbS)$_{1.18}$VS$_2$ crystal, where a 2H-type stacking domain structure is visible, with domain sizes up to tens of micrometers.}
  \label{fgr:fig_s6}
\end{figure}

Figure~\ref{fgr:fig_s6}, domains in a cleaved (PbS)$_{1.18}$VS$_2$ crystal are shown, where locally the layers are stacked in 2H-type stacking.

\section{XPS}

\begin{figure}
 \includegraphics[width=\columnwidth] {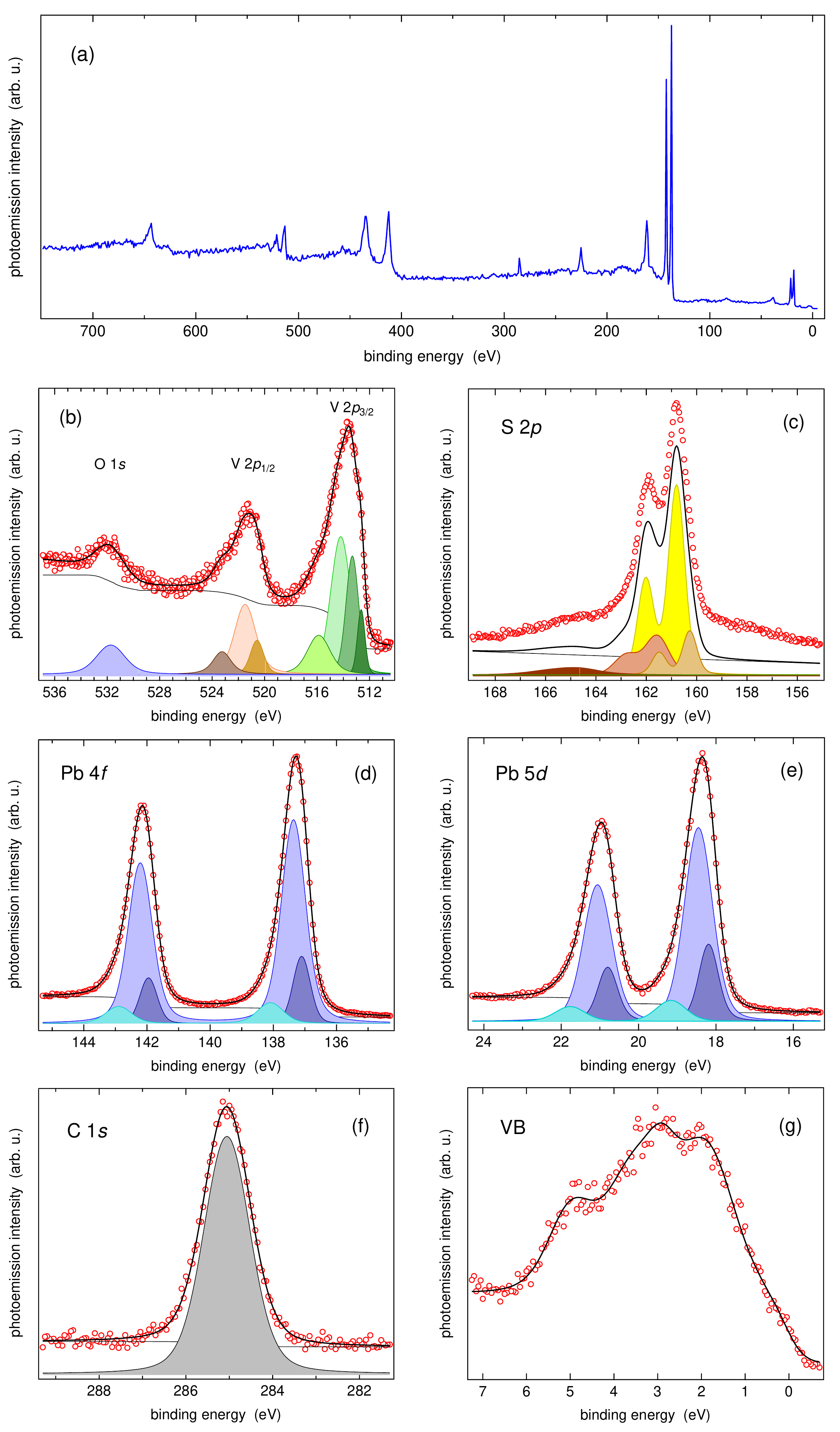}
  \caption{(a) XPS survey scan of the (PbS)$_{1.18}$VS$_2$ crystal. Panels (b--f) show V 2$p$ [with O 1$s$], S 2$p$, Pb 4$f$ and Pb 5$d$, and C 1$s$ core levels in detail. Fits are added assuming spin-orbit split Gaussian components and a Shirley background. For the fit of the V 2$p$ peak the overlapping, several eV wide, Pb 4$f$ loss peak was subtracted. 
  Panel (g) shows the valence band intensity close to the Fermi level. A smeared step-like Fermi edge is visible at zero binding energy.}
  \label{fgr:fig_s4}
\end{figure}

XPS core level (CL) spectroscopy allows to investigate the chemical state of elements in compounds.
Different coordination states of elements lead to energy shifted components in respective CLs spectra. 
In (PbS)$_{1.18}$VS$_2$, e.g., we expect two sulfur components reflecting its layer-dependent covalent chemical state in PbS and VS$_2$. Additional shifts can be induced for atoms located at the surface or at interfaces, in our case between weakly van der Waals bound PbS and VS$_2$ layers. \\ 
Due to the presence of interfaces between different layer materials, XPS data of MLC heterostructures is not straightforward to interpret. In first works, XPS data were compared with spectra of their individual layers. CL spectra were fit with a single peak, but the full width at half maximum (FWHM) of the MLC spectra was typically broader than those of individual layers~\cite{Ettema1993}. More recent, higher resolution spectra obtained of PbS and (PbS)$_{1.18}$NbS$_2$ showed that the FWHM does not only broaden, but also the asymmetry increases and more than one peak is needed to accurately fit the spectra~\cite{Brandt2003}.\\
The (PbS)$_{1.18}$VS$_2$ XPS data that we show in this section confirms some of the recent observations of multiple components in CLs in MLCs. Due to the complexity, we will, however, only mention some possible ways to interpret the CL shapes without a pretension to present an overall model.\\

In Fig.~\ref{fgr:fig_s4}, a XPS wide range scan [(a)] and detailed core levels of V 2$p$ [with O 1$s$], S 2$p$, Pb 4$f$ and Pb 5$d$, and C 1$s$ [(b) - (f)] are plotted for (PbS)$_{1.18}$VS$_2$. The CL spectra are shown together with fits based on standard Voigt components. During fitting, the number of Voigt components were kept as low as possible. Peak intensities and energy positions of single-Voigt C 1$s$ (peak energy 285.1~eV) and O 1$s$ (peak energy 531.8~eV) components correspond to spurious contamination like, e.g., adventitious carbon adsorbed on the surface and give evidence of clean surface properties after in situ cleave of our samples. 

We start our CL discussion with vanadium, which in VS$_2$ layers should be in a single chemical state sandwiched between sulfur atoms. The peak maxima of the V 2$p$ doublet in Fig.~\ref{fgr:fig_s4}(b) appear at about 512~eV (V 2$p_{3/2}$) and 520~eV (V 2$p_{1/2}$) consistent with recently chemical vapor deposition grown nanometer thick VS$_2$, VSe$_2$ and VTe$_2$ crystals~\cite{Hossain2018,Zhang2017,Hossain2021}. We comment here that for VO$_2$ the position of V 2$p_{3/2}$ would be expected at about 515~eV or 516~eV \cite{Silversmit2004}.

Our V 2$p_{3/2}$ spectra clearly exhibit an asymmetric line shape with three weak shoulders, resembling the recently reported spectral shapes of PbSe-VSe$_2$ heterostructures \cite{Goehler2022}. For 2$p$ levels of transition metals with a partially filled 3$d$ shell a multiplet structure is expected in XPS. In our case the main 2$p_{3/2}$ line can be fitted by three sharp Voigt components (512.5, 513,0 and 514~eV) and an additional broad component at 516~eV as shown in Fig.~\ref{fgr:fig_s4}(b).\\

Due to the different bonding conditions of S atoms in VS$_2$ and PbS layers, we expect at least two chemical states of sulfur. Indeed, the S 2$p$ peak in Fig.~\ref{fgr:fig_s4}(c) around 161~eV, is relatively broad and a composition of multiple peaks. The spin-orbit split S 2$p_{1/2}$ and S 2$p_{3/2}$ intensities each consist of a main component in the center and two minor ones at lower and higher binding energies. The latter is considerably broadened. For PbSe-VSe$_2$ heterostructures the XPS data of selenium shows a similar behavior\cite{Goehler2022}. Also here, Se 3$d$ in (PbSe)$_2$-(VSe$_2$)$_1$ can be fitted by two main components representing PbSe and VSe$_2$ plus an additional minor intensity at higher binding energy (not fitted in Ref. \cite{Goehler2022}).

For better understanding of the sulfur CL shapes we discuss the Pb CLs. The Pb 5$d$ (Pb 4$f$) CLs in Fig.~\ref{fgr:fig_s4}(d) and (e) are composed of two distinct Pb 4$d_{5/2}$ and Pb 4$d_{3/2}$ (Pb 4$f_{7/2}$ and Pb 4$f_{5/2}$) doublets. From the covalent bonding in PbS layers, we expect one single Pb chemical state, as observed in Pb 5$d$ CLs of cleaved PbS crystals measured with high resolution and low photon energies hv = 61~eV~\cite{LEIRO1998}. However, again we clearly observe an asymmetry in both Pb 5$d$ and Pb 4$f$ CL peaks, which can be decomposed into three Voigt components. The broader component at higher binding energies is needed to reproduce the asymmetric shape of the CLs (Comment: the asymmetry could be also covered by a Doniach-Sunjic line shape, which usually describe metallic systems.). For (PbSe)$_1$-(NbSe$_2$)$_2$ systems a similar asymmetry in Pb CLs was shown to be only present when PbSe layers are at an interface to NbSe$_2$ \cite{Goehler2018}. The effect was assigned to interface charge transfer of electrons from PbSe layers to NbSe$_2$, which shifts Pb CLs to higher binding energies. Consequently, Se 3$d$ CLs in PbSe layers are reported to shift by about 0.3~eV to lower energies compared to bulk PbSe.\\

Additional chemical shifts in CLs can stem from the influence of the surface properties, which can potentially depend on the presence of ferroelectric domains. At the boundary of a ferroelectric domain, the ferroelectric polarization induces a surface charge. A depolarization electric field is created by the surface charge. In ambient conditions, the presence of a thin water film plays an important role in the stabilization of the out-of-plane ferroelectric polarization via the screening of the depolarization field. In ultra-high vacuum (UHV) conditions, the role of water can be fulfilled by for example CO and CO$_2$ adsorbates, which are beside H$_2$ typical residual gases under UHV systems~\cite{Fong2006, Baniecki2009, Krug2010}.